\documentclass[preprint,11pt,oneside,leqno]{imsart}

\usepackage[margin=1in]{geometry}
\usepackage{amsthm,amsmath,amsfonts,amssymb}
\usepackage[authoryear]{natbib}
\usepackage{graphicx}
\usepackage{placeins}
\usepackage[table,dvipsnames]{xcolor}
\usepackage{tabularx,booktabs,array}
\usepackage{bm}
\usepackage{hyperref}
\makeatletter
\def\ps@imsheadings{%
  \def\@oddhead{\footnotesize\itshape\hfil\rightmark\hfil
    \llap{\normalfont\thepage}}%
  \let\@evenhead\@oddhead
  \let\@oddfoot\@empty
  \let\@evenfoot\@empty
}
\makeatother
\startlocaldefs

\newcommand{\ba}{ {\boldsymbol a} }

\newcommand{\bb}{ {\boldsymbol b} }
\newcommand{\bB}{ {\boldsymbol B} }

\newcommand{\bD}{ {\boldsymbol D} }
\newcommand{\be}{ {\boldsymbol e} }

\newcommand{\boldf}{ {\boldsymbol f} }

\newcommand{\bH}{ {\boldsymbol H} }
\newcommand{\bI}{ {\boldsymbol I} }

\newcommand{\bL}{ {\boldsymbol L} }
\newcommand{\bmM}{ {\boldsymbol m} }
\newcommand{\bM}{ {\boldsymbol M} }

\newcommand{\bN}{ {\boldsymbol N} }
\newcommand{\bo}{ {\boldsymbol o} }

\newcommand{\bp}{ {\boldsymbol p} }
\newcommand{\bP}{ {\boldsymbol P} }
\newcommand{\bq}{ {\boldsymbol q} }
\newcommand{\bQ}{ {\boldsymbol Q} }
\newcommand{\br}{ {\boldsymbol r} }
\newcommand{\bR}{ {\boldsymbol R} }
\newcommand{\bs}{ {\boldsymbol s} }

\newcommand{\bu}{ {\boldsymbol u} }

\newcommand{\bv}{ {\boldsymbol v} }
\newcommand{\bV}{ {\boldsymbol V} }
\newcommand{\bw}{ {\boldsymbol w} }

\newcommand{\bx}{ {\boldsymbol x} }

\newcommand{\by}{ {\boldsymbol y} }

\newcommand{\bz}{ {\boldsymbol z} }
\newcommand{\bZ}{ {\boldsymbol Z} }

\newcommand{\balpha}{ {\boldsymbol \alpha} }
\newcommand{\bbeta}{ {\boldsymbol \beta} }
\newcommand{\bgamma}{ {\boldsymbol \gamma} }
\newcommand{\bGamma}{ {\boldsymbol \Gamma} }

\newcommand{\bepsilon}{ {\boldsymbol \epsilon} }

\newcommand{\bkappa}{ {\boldsymbol \kappa} }
\newcommand{\blambda}{ {\boldsymbol \lambda} }
\newcommand{\bLambda}{ {\boldsymbol \Lambda} }

\newcommand{\bOmega}{ {\boldsymbol \Omega} }

\newcommand{\btheta}{ {\boldsymbol \theta} }

\newcommand{\bone}{ {\bf 1} }
\newcommand{\bzero}{ {\bf 0} }
\newcommand{\given}{\,|\,}
\newcommand{\tr}{^\top}
\newcommand{\eqnref}[1]{(\ref{#1})}
\newcommand{\eqnrefs}[2]{(\ref{#1}) and~(\ref{#2})}
\newcommand{\secref}[1]{Section~\ref{#1}}
\newcommand{\appref}[1]{Appendix~\ref{#1}}
\newcommand{\tabref}[1]{Table~\ref{#1}}

\newcommand{\figref}[1]{Figure~\ref{#1}}
\newcommand{\figrefs}[2]{Figures~\ref{#1} and~\ref{#2}}
\newcommand{\figrefrange}[2]{Figures~\ref{#1}--\ref{#2}}

\newcommand{\iid}{\overset{\mathrm{iid}}{\sim}}
\newcommand{\ind}{\overset{\mathrm{ind}}{\sim}}
\newcommand{\N}{\operatorname{N}}

\newcommand{\NB}{\operatorname{NB}}
\newcommand{\Ga}{\operatorname{Gamma}}
\newcommand{\Mult}{\operatorname{Multinomial}}
\newcommand{\Bin}{\operatorname{Binomial}}

\newcommand{\GP}{\operatorname{GP}}

\newcommand{\PG}{\operatorname{PG}}
\newcommand{\BAF}{\operatorname{BAF}}
\newcommand{\plotarea}{\mathrm{a}}
\newcommand{\E}{\operatorname{E}}
\newcommand{\Var}{\operatorname{Var}}

\newcommand{\diag}{\operatorname{diag}}

\newcommand{\IG}{\operatorname{IG}}
\newcommand{\Unif}{\operatorname{Unif}}

\endlocaldefs

\begin{document}

\begin{frontmatter}
\title{A Bayesian Magnitude-Composition Model for Tree-Size Distributions and Stand-Tables}
\runtitle{Magnitude-composition models for stand-tables}
\begin{aug}
\author[A]{\fnms{Andrew O.}~\snm{Finley}}
\author[B]{\fnms{Malcolm S.}~\snm{Itter}}
\author[C]{\fnms{Arthur}~\snm{Spitta}}
\author[C]{\fnms{Arne}~\snm{Nothdurft}}
\address[A]{Departments of Forestry and Statistics \& Probability, Michigan State University, USA}
\address[B]{Department of Environmental Conservation, University of Massachusetts Amherst, USA}
\address[C]{Department of Ecosystem Management, Climate and Biodiversity,\\
Institute of Forest Growth, BOKU University, Vienna, Austria}
\end{aug}

\begin{abstract}
Sparse forest inventories often provide too few plots to estimate species-specific tree-size distributions for individual stands. We develop a general Bayesian magnitude-composition model that separates overall abundance from its allocation among categories. A negative-binomial model describes total counts, while multinomial models describe how those counts are allocated among categories. In forest inventory, this construction allocates trees among species and then size classes within species, without imposing a parametric form on the tree-size distribution. Covariates inform both abundance and composition. Shared latent factors represent residual spatial and group-level dependence among categories using relatively few processes, while smoothing priors on regression coefficients borrow information across neighboring size classes. Nearest-neighbor Gaussian processes support efficient spatial computation. Joint posterior prediction and area-weighted aggregation yield stand-tables---tree densities by species and size class---and derived measures such as size-class diversity, with associated uncertainty. Simulations show small estimation errors and near-nominal predictive-interval coverage for forest and stand summaries. An application combining field plots and light detection and ranging (LiDAR) covariates illustrates estimation of species-specific distributions of diameter at breast height and size-class diversity in sampled and unsampled stands.
\end{abstract}

\end{frontmatter}

\section{Introduction}

Forest inventories often contain too few field plots to estimate species-specific tree-size distributions reliably for individual management units, or stands \citep{Brosofske2014forestinv,Henttonen2015plotdesign}. A stand-table summarizes these distributions as numbers of trees per unit area by species and size class \citep{Nanos2002diamdistr,Mehtaetalo2011diamdist,Nanos2017diamdistr}. These summaries inform silvicultural prescriptions, growth-and-yield projections, and habitat assessment, and provide indicators of structural diversity and biodiversity \citep{damato2011forest,storch2018structural,pretzsch2009growthmodels,degraaf1998birds}. Yet many stands contain few or no sampled plots, and individual species-by-size classes can be sparsely represented even where plots are available.

Remotely sensed covariates from light detection and ranging (LiDAR) offer a way to support estimation across these stands. LiDAR measurements describe canopy structure and can be available throughout a forest, linking conditions at sampled plots to those at unsampled locations \citep{White2016remotesens}. This setting motivates small area estimation (SAE), in which information from observations and auxiliary variables is combined to estimate quantities for domains with insufficient data for reliable direct estimation \citep{rao2015small}. The challenge here is to estimate complete species-by-size stand-tables, with uncertainty in both their individual entries and summaries derived from them.

A forest inventory at Weisse Krug in Germany illustrates this setting (\secref{sec:fixed_area_application}). It comprises 151 field plots across 249 stands, of which 139 contain no sampled plots, together with LiDAR covariates available across the forest. Tree size is recorded as diameter at breast height (DBH), measured at 1.30 m above ground. Our aim is to use these data to estimate species-by-DBH densities for sampled and unsampled stands and to describe variation in stand structure through DBH-class diversity. These targets require estimates of overall abundance as well as its allocation among species and DBH classes.

\subsection{Background and related work}

Traditional approaches to DBH distributions fit parametric families, such as the Weibull or Johnson's $S_B$, and relate their parameters or moments to stand characteristics \citep{bailey1973quantifying,hafley1977statistical,hyink1983generalized,poudel2013evaluation}. Such models provide compact descriptions, although their distributional forms can restrict the representation of irregular or multimodal size distributions. Mixtures and imputation offer greater flexibility, including transfer of observed tree lists or histograms to locations with similar auxiliary covariates \citep{liu2002finite,maltamo2009nonparametric}. LiDAR-assisted approaches also use distributional parameters, canopy metrics, and local calibration to predict DBH distributions \citep{gobakken2005weibull,hao2022uav,gallagherduval2023texture}. Individual-tree reconstruction provides another approach to estimating species and size distributions \citep{kostensalo2023recreating,kostensalo2026large}. Across these approaches, obtaining absolute species-by-DBH densities requires accounting for overall abundance, and uncertainty must be carried through any linked estimation steps.

Spatial hierarchical models provide a basis for joint estimation and uncertainty propagation. \citet{finley2014dynamic} modeled species-specific DBH counts using multivariate spatial Poisson regression with coefficients that evolve over diameter class, allowing dependence across locations and ordered classes. This model directly predicted stand-tables with uncertainty, although computation was demanding. \citet{nothdurft2024spatial} used Bayesian distributional regression with flexible additive predictors and spatial effects to map DBH distributions in alpine protection forests. Distributional regression can describe multiple features of a conditional distribution \citep{rigby2005gamlss,umlauf2018bamlss,kneib2023rage}; when the modeled quantities are relative DBH frequencies, an abundance estimate is also needed to obtain absolute stem densities.

Recent forestry SAE studies combine plot observations and auxiliary remote-sensing information to estimate quantities such as biomass, carbon, growing stock volume, basal area, and mean diameter in sparsely sampled domains \citep{Shannon2026gaswells,white2025saeczi,nothdurft2025smallarea}. Our setting extends this inferential task to a table of species-by-DBH densities within each stand. In a unit-level approach, relationships are modeled at the sampled plots and predictions are aggregated over the target stands. Dependence among categories and locations matters when forming these aggregates and calculating nonlinear summaries such as diversity.

\subsection{Proposed approach}

We develop a Bayesian magnitude-composition model that separates overall abundance from allocation among categories. A negative-binomial component models total plot counts, accounting for sampled area, and conditional multinomial components allocate those counts among species and then DBH classes within species. The construction gives overall abundance, species allocation, and size allocation within species separate regression and random-effect structures. A small number of shared latent factors captures residual spatial or group-level patterns in category allocation, while smoothing regularizes coefficients across ordered size classes. Together, these components permit joint estimation of species-specific size distributions without imposing a parametric density on tree diameters. Joint posterior prediction and area-weighted aggregation provide forest and stand estimates, with uncertainty propagated to the stand-tables and quantities derived from them.

We assess estimation and interval coverage using 200 simulated populations and compare candidate model specifications by cross-validation in the Weisse Krug analysis. The simulations show small average errors and coverage near nominal levels for forest and stand summaries. The inventory analysis describes differences in species-specific DBH distributions and stand-level size diversity, including estimates and uncertainty for stands without sampled plots. These analyses demonstrate how the framework supports inference on forest structure at scales finer than the inventory alone can reliably describe.

The following sections develop the model, computation, prediction, and assessment methods, followed by the simulation and inventory analyses and a discussion of their implications.

\section{Hierarchical magnitude-composition model}\label{sec:model}

To build model complexity incrementally, we start with a model for a one-dimensional (1-D) size-class stand-table whose entries are stem density or stem counts over a specified support. We then extend the model to estimate a two-dimensional (2-D) stand-table with size class on one margin and a second categorical variable defining the other margin. Data used to inform these estimates come from fixed-area forest inventory plots.

We present models with a rich set of terms, including fixed effects as well as unstructured and structured random effects. As described in \secref{sec:model_choice}, and demonstrated in \secref{sec:applications}, choosing an appropriately flexible model is part of applying these methods in practice. An extension to repeated inventories is discussed in \secref{sec:spatiotemporal_extensions}.

\subsection{Fixed-area plot sampling}

Consider a sample of $n$ fixed-area forest inventory plots. Plot $i$ is centered at coordinate vector $\bs_i$, has plot area $\plotarea_i$ shared by all categories, and records tree counts in $J$ categories ($i = 1,\ldots,n$). These categories may be tree-size classes, such as DBH classes. Let $y_{ij}$ denote the count in plot $i$ and category $j$, let $N_i=\sum_{j=1}^{J} y_{ij}$ denote the total count observed in plot $i$, and let $\by_i=(y_{i1},\ldots,y_{iJ})\tr$ denote the $J$-dimensional vector of observed counts in plot $i$.

We decompose the expected category-specific stem density into a scalar magnitude and a composition,
\begin{equation}
  \mu_{ij}=M_i p_{ij},
  \qquad
  p_{ij}\geq 0,
  \qquad
  \sum_{j=1}^{J} p_{ij}=1,
  \label{eq:mean-decomposition}
\end{equation}
where $\mu_{ij}$ is the expected category-specific stem density, $M_i$ is the unknown expected overall stem density (summed across categories), and the vector $\bp_i=(p_{i1},\ldots,p_{iJ})\tr$ contains category probabilities that allocate this density among categories. The magnitude likelihood for the total count is
\begin{equation}
  N_i \given M_i,r_M
  \sim
  \NB(\plotarea_i M_i,r_M),
  \label{eq:total-count}
\end{equation}
where the negative-binomial (NB) distribution is parameterized by the mean total count $\plotarea_iM_i$ for plot $i$ and the size parameter $r_M>0$. Thus,
\begin{equation}
  \E(N_i\given M_i,r_M)=\plotarea_i M_i,
  \qquad
  \Var(N_i\given M_i,r_M)=
  \plotarea_i M_i
  +
  \frac{(\plotarea_i M_i)^2}{r_M}.
  \label{eq:nb-mean-variance}
\end{equation}
The Poisson total-count model is the limiting case as $r_M\rightarrow\infty$. The allocation likelihood is
\begin{equation}
  \by_i \given N_i,\bp_i \sim \Mult(N_i,\bp_i).
  \label{eq:composition-count}
\end{equation}
Together, \eqnrefs{eq:total-count}{eq:composition-count} imply
\begin{equation}
  \E(y_{ij}\given M_i,\bp_i,r_M)=\plotarea_i M_i p_{ij}=\plotarea_i \mu_{ij},
  \label{eq:fixed-area-cell-mean}
\end{equation}
This identity links the per-unit-area target to the plot-sampling scale, with $\plotarea_i\mu_{ij}$ giving the expected count observed on plot $i$ in category $j$. The factorization separates two inferential targets, overall stem density and its allocation among size classes. We assume observations from different plots are independent conditional on the model parameters and latent effects.

The magnitude regression model is specified on the log scale,

\begin{equation}
  \log M_i =
  \bz_i\tr\bbeta + \delta_{g(i)} + v(\bs_i),
  \label{eq:magnitude-model}
\end{equation}
where $\bz_i$ contains plot-level covariates, such as LiDAR canopy height metrics, and a leading one for the intercept, $\bbeta$ is the corresponding vector of regression coefficients, $\delta_{g(i)}$ is a random effect for group $g(i)$ (e.g., stand or stratum) with $\delta_g\given\sigma_\delta^2\iid\N(0,\sigma_\delta^2)$, and $v(\bs_i)$ is a spatial random effect. The log scale ensures $M_i$ is positive. The specification in \eqnref{eq:magnitude-model} shows one group-level and one spatial term to keep notation readable; additional group or spatial process terms can enter additively. We write the spatial effect as a Gaussian process (GP),
\begin{equation}
  v(\cdot) \sim \GP\{0,\sigma_v^2 \rho_v(\cdot,\cdot;\phi_v)\},
  \label{eq:magnitude-gp}
\end{equation}
where $\rho_v$ is a valid spatial correlation function. In our applications, we use the exponential form $\rho_v(\bs,\bs';\phi_v)=\exp\{-\phi_v\|\bs-\bs'\|\}$, where $\|\bs-\bs'\|$ is the Euclidean distance between locations $\bs$ and $\bs'$ and $\phi_v>0$ is the spatial decay parameter \citep{banerjee2014hierarchical}.

Although \eqnref{eq:magnitude-model} is richly specified, applied analyses can remove terms when simpler submodels are preferred or when the data do not support all terms. Or, additional terms can be added to capture other grouping structures or accommodate covariate effects that vary by group through a random slope or vary continuously over space through a spatially-varying coefficient process \citep{gelfand2003spatial,gelfand2004nonstationary}.

For fixed-effect magnitude models with no random effects, estimating $r_M$ provides a direct way to represent plot-level overdispersion in total abundance. When magnitude random effects are included, latent heterogeneity in $M_i$ also produces extra-Poisson variation. To reduce confounding between count-level dispersion and process variation, we fix $r_M$ at a large, finite value in these models. We retain the negative-binomial likelihood for computational convenience and use the same sampling framework across magnitude model specifications (\secref{sec:computation}).

\subsection{Composition model}

The multinomial composition in \eqnref{eq:composition-count} is written using a con\-tinuation-ratio representation \citep{tutz1991sequential,agresti2013categorical}. Given a fixed category order, this is an exact reparameterization of the multinomial allocation. Let $R_{ij}$ denote the number of trees remaining to be allocated before category $j$. For the first category, all $N_i$ trees are available, so $R_{i1}=N_i$. For subsequent categories, subtracting the counts already allocated gives
\begin{equation}
  R_{ij}=N_i-\sum_{h<j} y_{ih},
  \qquad j=1,\ldots,J-1.
  \label{eq:remaining}
\end{equation}
The count allocated to category $j$ then follows
\begin{equation}
  y_{ij}\given R_{ij},\pi_{ij}
  \sim
  \Bin(R_{ij},\pi_{ij}),
  \qquad j=1,\ldots,J-1,
  \label{eq:sequential-binomial}
\end{equation}
where $\pi_{ij}$ is the conditional probability that a tree belongs to category $j$, given that it does not belong to an earlier category. The final category receives the remaining trees, $y_{iJ}=R_{i,J-1}-y_{i,J-1}$.

The corresponding multinomial category probabilities are
\begin{equation}
  p_{ij}
  =
  \pi_{ij}\prod_{h<j}(1-\pi_{ih}),
  \qquad j=1,\ldots,J-1,
  \label{eq:stick-transform}
\end{equation}
and
\begin{equation}
  p_{iJ}
  =
  \prod_{h<J}(1-\pi_{ih}).
  \label{eq:stick-final}
\end{equation}

The composition linear predictor is
\begin{equation}
  \begin{aligned}
  \psi_{ij}
  &=
  \alpha_{0j} + \bx_i\tr\balpha_j +
  \bgamma_j\tr\bu_{h(i)} +
  \blambda_j\tr \bw(\bs_i),\\
  \pi_{ij}
  &=
  \Phi(\psi_{ij}),
  \qquad j=1,\ldots,J-1,
  \end{aligned}
  \label{eq:composition-model}
\end{equation}
where $\Phi(\cdot)$ is the standard normal distribution function. Here $\bx_i$ is a vector of plot-level covariates, $\alpha_{0j}$ is a category-specific intercept, $\balpha_j$ is a vector of category-specific composition regression coefficients, $\bgamma_j$ is a loading vector for unstructured factor scores $\bu_{h(i)}$, and $\blambda_j$ is a loading vector for spatially structured factor scores $\bw(\bs_i)$. Unlike in the magnitude model \eqnref{eq:magnitude-model}, here we make the model intercept $\alpha_{0j}$ explicit because it can be assigned a cross-category smoothing prior, as described in \secref{sec:prior_specification}. Let $\bGamma=(\bgamma_1,\ldots,\bgamma_{J-1})\tr$ and $\bLambda=(\blambda_1,\ldots,\blambda_{J-1})\tr$ collect the unstructured and spatial loading vectors over the continuation-ratio linear predictors. The products $\bGamma\bu_{h(i)}$ and $\bLambda\bw(\bs_i)$ give low-rank residual dependence across those linear predictors, allowing categories to share group-level or spatial patterns without assigning a separate random effect to every category. For a grouping variable $h(i)$, the unstructured factor scores satisfy $\bu_h\iid\N(\bzero,\bI)$, with scale carried by $\bGamma$. The latent spatial factor process is $\bw(\bs)=(w_1(\bs),\ldots,w_K(\bs))\tr$, with
\begin{equation}
  w_k(\cdot) \ind
  \GP\{0,\rho_{w,k}(\cdot,\cdot;\phi_{w,k})\},
  \qquad k=1,\ldots,K.
  \label{eq:factor-gp}
\end{equation}
In our applications, $\rho_{w,k}$ has the same exponential form as $\rho_v$, with factor-specific decay parameter $\phi_{w,k}>0$.

For parameter identifiability, each unstructured or spatial factor term has unit-variance factors and a $(J-1)\times K$ loading matrix, with $K\leq J-1$. The number of factors $K$ may differ among terms. We select and order $K$ response categories as anchors and constrain their loading-row submatrix to have zeros above the diagonal and strictly positive diagonal entries that are estimated \citep{lopes2004bayesian,taylorrodriguez2019spatial}. The anchors establish factor labels and signs and should be well represented in the data to support estimation. All other loadings are unrestricted.

As with the magnitude model, \eqnref{eq:composition-model} is richly specified and simpler submodels may be preferred. Richer specifications might include additional grouping effects and group- or spatially-varying coefficients. In the composition model, these extensions are written in factor form, so a small number of latent group-level or continuous spatial coefficient patterns are mapped to many continuation-ratio linear predictors through loading matrices.

\subsection{Ordered and unordered categories}

Categorical variables used to define a stand-table's margins can be ordered or unordered, and this distinction has model specification and estimation implications. Tree-size classes are ordered: adjacent classes represent adjacent portions of the size distribution. It is therefore reasonable to pool information across neighboring size classes, allowing the category-specific intercepts or regression coefficients to vary smoothly over the ordered categories.

Nominal categorical variables such as species, forest types, and ownership have no natural ordering. A continuation-ratio model still requires an ordering, but for nominal categories that ordering is part of the probability factorization rather than a structure for borrowing information across proximate categories. Different orderings define different sequences of conditional binomial splits, so the nominal order should be reported as part of the model specification. Because nominal categories do not have a natural adjacency, we do not impose smoothing across them. Independent or exchangeable priors are more appropriate for unordered categories because they allow shrinkage without asserting an artificial neighborhood.

\subsection{Conditional two-way stand-tables}

We extend the 1-D stand-table to a 2-D stand-table using a conditional construction. Let $A$ index a leading categorical variable with categories $a=1,\ldots,J_A$, and let $B$ index a second categorical variable with categories $b=1,\ldots,J_B$. In the species-by-DBH setting, $A$ indexes species group and $B$ indexes DBH class. We write the joint probability that a tree in plot $i$ belongs to category $a$ of $A$ and category $b$ of $B$ as
\begin{equation}
  p_{iab}=q_{ia}r_{ib\given a},
  \label{eq:two-way-factorization}
\end{equation}
where $q_{ia}$ is the plot-specific probability for category $a$ of $A$, and $r_{ib\given a}$ is the plot-specific probability for category $b$ of $B$ conditional on $A=a$; equivalently, $q_{ia}=P(A=a)$ and $r_{ib\given a}=P(B=b\given A=a)$. Species groups are unordered and DBH classes are ordered, so smoothing can be imposed across DBH classes within species groups but not across species groups themselves.

The corresponding composition likelihood is represented by two allocation blocks. Let $y_{ia+}=\sum_{b=1}^{J_B}y_{iab}$, $\by_{iA+}=(y_{i1+},\ldots,y_{iJ_A+})\tr$, and $\by_{iaB}=(y_{ia1},\ldots,y_{iaJ_B})\tr$. The leading block allocates the total count among the categories of $A$,
\begin{equation}
  \by_{iA+}\given N_i,\bq_i
  \sim
  \Mult(N_i,\bq_i),
  \label{eq:leading-margin}
\end{equation}
where $\bq_i=(q_{i1},\ldots,q_{iJ_A})\tr$. The conditional block allocates the parent count among the categories of $B$ within each $A=a$,
\begin{equation}
  \by_{iaB}\given y_{ia+},\br_{i\cdot\given a}
  \sim
  \Mult(y_{ia+},\br_{i\cdot\given a}),
  \qquad a=1,\ldots,J_A,
  \label{eq:child-conditional}
\end{equation}
where $\br_{i\cdot\given a}=(r_{i1\given a},\ldots,r_{iJ_B\given a})\tr$.
Together with the total-count magnitude model, these allocation blocks imply cell means $\plotarea_i M_i q_{ia}r_{ib\given a}$; the total count is modeled once and then allocated through the leading and conditional composition blocks.

Each allocation block is again written using continuation-ratio probabilities. Let $\pi^{(A)}_{ia}$ denote the conditional probability for parent category $a$, given membership has not been assigned to an earlier parent category. Its linear predictor is
\begin{equation}
  \begin{aligned}
  \psi^{(A)}_{ia}
  &=
  \alpha^{(A)}_{0a} + \bx_i^{(A)\tr}\balpha^{(A)}_a +
  \left(\bgamma^{(A)}_a\right)\tr\bu^{(A)}_{h(i)} +
  \left(\blambda^{(A)}_a\right)\tr\bw^{(A)}(\bs_i),\\
  \pi^{(A)}_{ia}
  &=
  \Phi(\psi^{(A)}_{ia}),
  \qquad a=1,\ldots,J_A-1.
  \end{aligned}
  \label{eq:leading-composition-model}
\end{equation}
For the conditional block, let $\pi^{(B\given A)}_{iab}$ denote the conditional probability for child category $b$ within parent category $a$, given membership has not been assigned to an earlier child category. Its linear predictor is
\begin{equation}
  \begin{aligned}
    \psi^{(B\given A)}_{iab}
    &=
    \alpha^{(B\given A)}_{0ab} +
    \bx_i^{(B\given A)\tr}\balpha^{(B\given A)}_{ab}
    +
    \left(\bgamma^{(B\given A)}_{ab}\right)\tr\bu^{(B\given A)}_{h(i)}
    +
    \left(\blambda^{(B\given A)}_{ab}\right)\tr\bw^{(B\given A)}(\bs_i),\\
    \pi^{(B\given A)}_{iab}
    &=
    \Phi(\psi^{(B\given A)}_{iab}),\\
    &\hspace{1.5em} a=1,\ldots,J_A,\quad b=1,\ldots,J_B-1,
  \end{aligned}
  \label{eq:conditional-composition-model}
\end{equation}
where $\alpha^{(B\given A)}_{0ab}$ is the intercept for child category $b$ within parent category $a$, $\bx_i^{(B\given A)}$ is a vector of plot-level covariates, and $\balpha^{(B\given A)}_{ab}$ is the corresponding vector of regression coefficients. The vectors $\bgamma^{(B\given A)}_{ab}$ and $\blambda^{(B\given A)}_{ab}$ contain loadings for the unstructured factor scores $\bu^{(B\given A)}_{h(i)}$ and spatial factor scores $\bw^{(B\given A)}(\bs_i)$, respectively. Thus each parent category has its own set of conditional-allocation coefficients and loadings.

The covariate vectors $\bx_i^{(A)}$ and $\bx_i^{(B\given A)}$ need not contain the same variables. For example, covariates for allocation among species groups may differ from those describing the DBH distribution within species groups.

The conditional DBH model uses a common set of spatial or group-level factors across species. These factors can represent residual patterns associated with unmeasured site conditions, past management, or disturbance history. Each species has its own loadings that determine how these patterns affect its DBH distribution. A common pattern could therefore shift one species toward smaller DBH classes and another toward larger classes. This represents dependence among species' DBH distributions using relatively few factors.

Each conditional-block loading matrix has $J_A(J_B-1)$ rows, one for each combination of parent category and child continuation-ratio split, and each factor term selects its anchors from these rows.

\subsection{Prior specification}\label{sec:prior_specification}

Regression coefficients in the magnitude and composition models receive independent normal priors,
\begin{equation}
  \bbeta \sim \N(\bzero,\sigma_\beta^2\bI),
  \qquad
  \alpha_{0j} \ind \N(0,\sigma_{\alpha,0}^2),
  \qquad
  \balpha_j \ind \N(\bzero,\sigma_\alpha^2\bI),
  \label{eq:coefficient-priors}
\end{equation}
with fixed prior variances chosen to be weakly informative on the respective link scales. For ordered composition blocks, the independent priors can be replaced by a normal smoothing prior. Let $\btheta$ collect the intercepts or one covariate's coefficients across the ordered categories. We assign
\begin{equation}
  \btheta \sim
  \N\left\{
    \bzero,
    \left(\tau^{-2}\bD_d\tr\bD_d+\bV^{-1}\right)^{-1}
  \right\},
  \label{eq:smoothing-prior}
\end{equation}
where $\bD_d$ forms differences of order $d$. The applications use $d=2$, giving second-order random-walk (RW2) smoothing. The fixed variance $\tau^2$ controls the strength of smoothing, and $\bV$ is the corresponding covariance from \eqnref{eq:coefficient-priors}, which makes the prior proper. This prior is applied separately to each coefficient vector, using its corresponding smoothing variance and prior covariance. In conditional models, smoothing is applied separately within each parent category.

The magnitude group and spatial variances receive independent inverse-gamma priors,
\begin{equation}
  \sigma_\delta^2 \sim \IG(a_\delta,b_\delta),
  \qquad
  \sigma_v^2 \sim \IG(a_\sigma,b_\sigma),
  \label{eq:variance-priors}
\end{equation}
where $a_\cdot$ and $b_\cdot$ are the shape and scale, respectively. Each additional magnitude random-effect term has its own variance and corresponding prior.

The remaining scalar parameters are the negative-binomial size and spatial decays, with priors
\begin{equation}
  r_M \sim \Ga(a_r,b_r),
  \qquad
  \phi \sim \Unif(\phi_L,\phi_U),
  \label{eq:scalar-parameter-priors}
\end{equation}
where $a_r$ and $b_r$ are the shape and rate, respectively. The size prior applies when $r_M$ is estimated; models with magnitude random effects fix $r_M$ at a prespecified finite value. The uniform prior applies independently to each estimated magnitude or composition decay, with positive bounds chosen to represent plausible spatial scales.

For composition factor terms, unrestricted loadings receive independent normal priors,
\begin{equation}
  \gamma_{jk}\ind\N(0,c_\gamma),
  \qquad
  \lambda_{jk}\ind\N(0,c_\lambda).
  \label{eq:loading-priors}
\end{equation}
Anchor diagonals receive the corresponding independent normal priors restricted to positive values, giving half-normal priors. The fixed normal-prior variances $c_\gamma$ and $c_\lambda$ may differ among factor terms. Entries constrained to zero remain fixed.

\section{Posterior computation and spatial approximation}\label{sec:computation}

Inference uses Markov chain Monte Carlo (MCMC) samples from the posterior defined by the model in \secref{sec:model} and the priors in \secref{sec:prior_specification}. Posterior draws retained after discarding burn-in and applying any thinning provide parameter and latent-process summaries, fitted values, and predictions as described in \secref{sec:prediction_aggregation}. We use multichain traceplots and the potential scale reduction statistic $\widehat R$ to assess convergence. Effective sample size (ESS) summarizes sampling efficiency in the presence of autocorrelation \citep{gelman2013bayesian}.

The sampler uses P\'olya--Gamma augmentation for the negative-binomial magnitude likelihood \citep{polson2013polya,hamura2025robust} and latent-normal augmentation for the probit composition likelihood \citep{albert1995bayesian,rouder2005introduction}. Conditional on the augmented variables, regression coefficients, factor scores, and latent effects have normal Gibbs updates. Loadings not fixed at zero are sampled from normal full conditional distributions, truncated to positive values for anchor diagonals. Magnitude process variances have inverse-gamma Gibbs updates; estimated spatial decays and negative-binomial size use Metropolis updates. Supplementary Section~\ref{app:sampler} gives the full conditional updates, Metropolis steps, and their implementation using dense and sparse Cholesky solves.

For the spatial fits, we replace the GPs in \secref{sec:model} with nearest-neighbor Gaussian processes (NNGPs) to reduce computational cost \citep{datta2016hierarchical}. These approximations use the specified exponential covariance functions as their parent covariances. For fitting, we order locations by the first spatial coordinate, using the second to break ties, and use up to $m=15$ nearest earlier locations as neighbors. Spatial predictions are generated jointly over the target grid, preserving dependence for stand and forest aggregation.

\section{Fitted values, prediction, and aggregation}\label{sec:prediction_aggregation}

Fitted values describe expected quantities at the observed plots, whereas predictions concern locations not used in fitting. Both are obtained from posterior draws of the model parameters and latent effects. Let $\ell$ index retained posterior draws. For the fixed-area model, evaluating the magnitude and composition models at observed plot $i$ gives the fitted category-specific density $\mu_{ij}^{(\ell)}=M_i^{(\ell)}p_{ij}^{(\ell)}$ and expected count $\plotarea_i\mu_{ij}^{(\ell)}$. These calculations use the sampled group and spatial effects at the observed plots. Posterior means provide fitted point estimates, and posterior quantiles describe uncertainty about the expected quantities.

At a new location $\bs_0$, we evaluate the regression terms using that location's covariates and draw any unobserved latent effects. Spatial effects are drawn jointly across prediction locations, conditional on the sampled effects at observed locations and the covariance parameters. If group effects are included, an existing group's sampled effect is retained, while a new group's effect is drawn from its model distribution conditional on the posterior parameters. Locations in the same group share its effect. Substitution into the magnitude and composition models gives $M^{(\ell)}(\bs_0)$, $\bp^{(\ell)}(\bs_0)$, and the expected category-specific density
\begin{equation}
  \mu_j^{(\ell)}(\bs_0)=M^{(\ell)}(\bs_0)p_j^{(\ell)}(\bs_0).
  \label{eq:prediction-cell}
\end{equation}
For two-way stand-tables, the category probabilities are constructed from the leading and conditional allocations using \eqnref{eq:two-way-factorization}. Conditional on these quantities and the size parameter $r_M^{(\ell)}$, we predict a count for a unit of area $\plotarea(\bs_0)$ by drawing
\begin{equation}
  \begin{aligned}
    \widetilde N^{\,(\ell)}(\bs_0)
    &\sim \NB\{\plotarea(\bs_0)M^{(\ell)}(\bs_0),r_M^{(\ell)}\},\\
    \widetilde{\by}^{\,(\ell)}(\bs_0)
    &\sim \Mult\{\widetilde N^{\,(\ell)}(\bs_0),\bp^{(\ell)}(\bs_0)\}.
  \end{aligned}
  \label{eq:predictive-counts}
\end{equation}
Here tildes denote posterior predictive draws; if size is fixed, $r_M^{(\ell)}=r_M$. The resulting stem density is $\widetilde d_j^{\,(\ell)}(\bs_0)=\widetilde y_j^{\,(\ell)}(\bs_0)/\plotarea(\bs_0)$. These draws include count variation as well as uncertainty in parameters and latent effects. Conditional on the parameters and latent effects, the mean of this density is $\mu_j^{(\ell)}(\bs_0)$ in \eqnref{eq:prediction-cell}.

Stand and forest summaries are obtained by aggregating joint posterior predictive draws over a grid of locations representing the target area \citep{finley2024models}. Predicted counts are converted to densities and weighted by the land area represented by each prediction location. This provides a grid approximation to the desired summaries, with uncertainty propagated by performing the aggregation separately within each posterior draw \citep{kerman2007manipulating}.

Specifically, for a target area $S$, let $\mathcal{S}_0$ be a finite set of prediction locations representing portions of $S$. Each location $\bs_0\in\mathcal{S}_0$ represents land area $A(\bs_0)$ within $S$, with these areas summing to the area of $S$. This area supplies the aggregation weight and can differ from the area $\plotarea(\bs_0)$ of the prediction unit. If $j$ indexes the full species-by-DBH table, the grid approximation to the category total for draw $\ell$ is
\begin{equation}
  \widetilde T_{Sj}^{\,(\ell)}=\sum_{\bs_0\in\mathcal{S}_0} A(\bs_0)\widetilde d_j^{\,(\ell)}(\bs_0),
  \label{eq:stand-total}
\end{equation}
and the corresponding category density is
\begin{equation}
  \widetilde d_{Sj}^{\,(\ell)}
  =
  \frac{\widetilde T_{Sj}^{\,(\ell)}}
       {\sum_{\bs_0\in\mathcal{S}_0} A(\bs_0)}.
  \label{eq:stand-density}
\end{equation}
Changing $S$ gives summaries for individual stands or the entire forest. This aggregation of local predictions provides the unit-level small area estimates used in the applications.

The full species-by-DBH table also supplies overall stem density and the species and DBH marginal distributions. Using the species and DBH-class indices $a$ and $b$, respectively, we write the cell totals and densities in \eqnrefs{eq:stand-total}{eq:stand-density} as $\widetilde T_{Sab}^{\,(\ell)}$ and $\widetilde d_{Sab}^{\,(\ell)}$. Within each posterior draw,
\begin{equation}
  \widetilde d_S^{\,(\ell)}
  =\sum_{a=1}^{J_A}\sum_{b=1}^{J_B}\widetilde d_{Sab}^{\,(\ell)},
  \qquad
  \widetilde d_{Sa+}^{\,(\ell)}
  =\sum_{b=1}^{J_B}\widetilde d_{Sab}^{\,(\ell)},
  \qquad
  \widetilde d_{S+b}^{\,(\ell)}
  =\sum_{a=1}^{J_A}\widetilde d_{Sab}^{\,(\ell)}.
  \label{eq:stand-derived-summaries}
\end{equation}
Here $+$ denotes summation over the corresponding category index. The same sums can be applied to predicted totals $\widetilde T_{Sab}^{\,(\ell)}$. Derived quantities, such as species diversity or the size-class diversity metric introduced in \secref{sec:dbh_diversity}, are also calculated within each draw before computing posterior summaries.

We summarize each aggregated quantity by its posterior predictive mean and a central 95\% credible interval under its posterior predictive distribution, defined by the 2.5th and 97.5th percentiles of the draws. We refer to these as posterior predictive intervals. The draw-wise calculations retain dependence among categories and prediction locations through the shared parameters and latent effects, and include negative-binomial and multinomial count variation.

\section{Model choice}\label{sec:model_choice}

We use leave-one-plot-out cross-validation to compare candidate models. Each plot is held out in turn, the model is fit to the remaining plots, and species-by-DBH densities are predicted at the held-out plot using its covariates and coordinates. Assessment focuses on predictive performance across the full species-by-DBH table.

Point predictions are posterior predictive means, calculated by averaging expected densities over posterior draws. Predictive densities and intervals are obtained using \eqnref{eq:predictive-counts}, with each held-out plot's sampled area. These draws include count variation and uncertainty in parameters and latent effects.

Our primary criterion is the continuous ranked probability score (CRPS), averaged over held-out plots and species-by-DBH cells. We favor CRPS because it evaluates the entire predictive distribution for each cell, including its uncertainty. It is a strictly proper scoring rule, expressed in the same units as density, with lower scores indicating better prediction \citep{gneiting2007strictly}. Root mean squared prediction error (RMSPE) evaluates errors in posterior mean predictions and gives greater weight to large errors. We retain RMSPE as a complementary measure of point-prediction accuracy.

Mean error and relative mean error summarize signed errors and indicate overall over- or under-prediction. Coverage reports the proportion of held-out observations within central 95\% posterior predictive intervals, while mean interval width summarizes their precision. Coverage and width are interpreted together, accounting for the discreteness of the observations. Definitions of these metrics are given in Supplementary Section~\ref{sec:cv_metrics}.

Final selection also considers the other assessment metrics, model complexity, and MCMC diagnostics. When predictive performance is similar, we favor a simpler model with better agreement between MCMC chains.

\section{Implementation}\label{sec:implementation}

The modeling framework and sampling algorithms (Supplementary Section~\ref{app:sampler}) are implemented in the \textsf{magComp} package \citep{finley2026magcomp} for R \citep{RCoreTeam2024}. Computational routines are written in C++. Dense matrix operations call the Basic Linear Algebra Subprograms (BLAS; \citealp{blackford2002blas}) and Linear Algebra PACKage (LAPACK; \citealp{anderson1999lapack}) through their Fortran interfaces, using the libraries linked to R. This permits use of optimized, multithreaded BLAS implementations such as OpenBLAS \citep{zhang2012openblas}. Sparse Cholesky factorizations and solves use CHOLMOD \citep{chen2008cholmod} through the Matrix package \citep{bates2026matrix}.

Source code and vignettes are available from the \textsf{magComp} package's GitHub repository. The vignettes use simulated data and worked examples to explain model components, fitting, prediction, and interpretation of results.

\section{Applications}\label{sec:applications}

We evaluate the modeling framework through a simulation study and an analysis of forest inventory data collected at Weisse Krug in Germany. The inventory and accompanying remotely sensed data are described in \secref{sec:fixed_area_application}. These data also provide the setting for the simulated populations. The simulation study assesses estimation performance and interval coverage for model parameters, forest and stand densities, and stand diversity across populations with known values. The Weisse Krug analysis then illustrates model choice and estimation of species-specific DBH distributions and stand diversity.

\subsection{Weisse Krug data}
\label{sec:fixed_area_application}

The motivating data come from a forest inventory conducted at Weisse Krug in Mecklenburg-Vorpommern, Germany, between the villages of Warin and Blankenberg. The mapped forest comprises 249 stands totaling 513.1 ha. Stand area ranges from 0.06 to 17.77 ha, with a mean of 2.06 ha and a median of 1.51 ha. The forest is mixed-species, dominated by European beech (\textit{Fagus sylvatica}) and Scots pine (\textit{Pinus sylvestris}).

The field inventory followed a systematic grid with approximately 183 m spacing, a randomized offset, and a randomized rotation. The resulting sampling grid contained 165 potential plot locations; 14 were inaccessible, leaving 151 measured plots. Measurements were made on 20 m radius circular plots in March 2026. The analysis includes trees at least 10 cm DBH. For plots intersecting the forest boundary, tree counts and sampled areas include only the forested portion of the plot. Sampled areas range from 0.0165 to 0.1257 ha, with a mean of 0.1200 ha. The resulting data contain 6,607 trees. Mean plot density, calculated by dividing each plot's count by its sampled area and averaging across plots, is 363.4 trees/ha.

Trees are assigned to 12 species groups and fifteen DBH classes: fourteen 5 cm intervals from 10--15 to 75--80 cm, followed by an open-ended class of $\geq$80 cm. The species groups are European beech (\textit{Fagus sylvatica}), Scots pine (\textit{Pinus sylvestris}), oak (\textit{Quercus} spp.), black alder (\textit{Alnus glutinosa}), Norway spruce (\textit{Picea abies}), Douglas fir (\textit{Pseudotsuga menziesii}), larch (\textit{Larix} spp.), silver birch (\textit{Betula pendula}), grand fir (\textit{Abies grandis}), maple (\textit{Acer} spp.), other broadleaves, and willow (\textit{Salix} spp.). The species-by-DBH table therefore contains 180 cells ($12\times15$). \tabref{tab:real_inventory_species} summarizes species abundance and occupancy.

\begin{table}[!ht]
\centering
\small
\setlength{\tabcolsep}{3pt}
\caption{Observed abundance and occupancy by species group in the Weisse Krug inventory. Mean density averages species-specific tree counts divided by sampled plot area across all 151 plots, including zeros where the species is absent. Occupied plots, DBH classes, and plot $\times$ DBH cells contain at least one tree of the species group. Each species group has 151 possible plots, 15 possible DBH classes, and $151\times15=2{,}265$ possible plot $\times$ DBH cells.}
\label{tab:real_inventory_species}
\begin{tabular*}{\textwidth}{@{\extracolsep{\fill}}lrrrrr@{}}
\toprule
Species group & \multicolumn{1}{c}{Measured} & \multicolumn{1}{c}{Mean density} & \multicolumn{1}{c}{Occupied} & \multicolumn{1}{c}{Occupied} & \multicolumn{1}{c@{}}{Occupied} \\
 & \multicolumn{1}{c}{trees} & \multicolumn{1}{c}{(trees/ha)} & \multicolumn{1}{c}{plots} & \multicolumn{1}{c}{DBH classes} & \multicolumn{1}{c@{}}{plot $\times$ DBH cells} \\
\midrule
European beech & 2160 & 116.4 & 117 & 15 & 673 \\
Scots pine & 1343 & 72.1 & 63 & 13 & 267 \\
Oak & 663 & 36.6 & 66 & 15 & 189 \\
Black alder & 518 & 30.7 & 28 & 13 & 134 \\
Norway spruce & 509 & 28.6 & 41 & 13 & 145 \\
Douglas fir & 493 & 28.0 & 22 & 13 & 128 \\
Larch & 230 & 12.9 & 25 & 11 & 88 \\
Silver birch & 194 & 11.1 & 47 & 11 & 109 \\
Grand fir & 188 & 9.9 & 2 & 7 & 14 \\
Maple & 150 & 8.1 & 23 & 10 & 55 \\
Other broadleaves & 126 & 7.0 & 35 & 9 & 72 \\
Willow & 33 & 2.0 & 3 & 11 & 14 \\
\bottomrule
\end{tabular*}
\end{table}

\begin{figure}[!ht]
\centering
\includegraphics[width=0.78\textwidth]{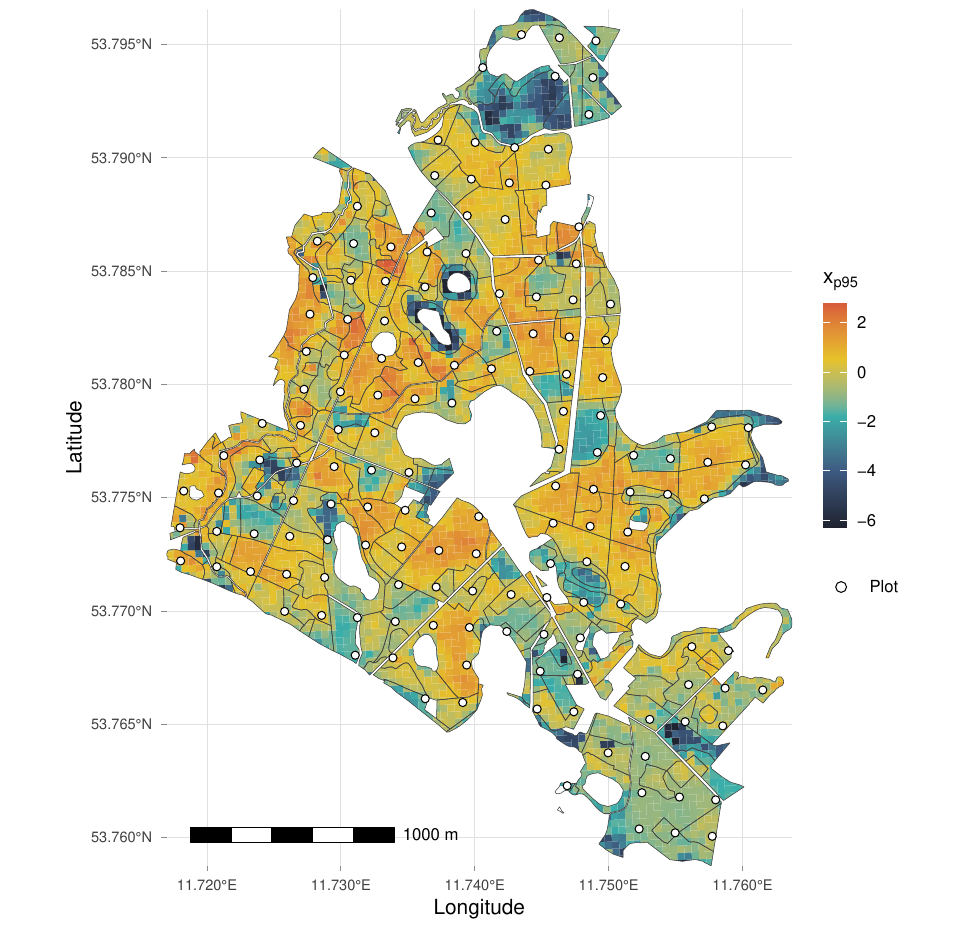}
\caption{Weisse Krug forest stands and field inventory. The surface shows the standardized 95th-percentile canopy-height covariate $x_{\mathrm{p95}}$ over the prediction cells, gray lines outline the 249 forest stands, and white circles mark the 151 field plots. Covariate values are standardized using the field plots' mean and standard deviation.}
\label{fig:weissekrug_data_map}
\end{figure}

The auxiliary data are derived from uncrewed aerial vehicle (UAV)-borne laser scanning acquired in March 2026, close to the field inventory. They comprise a 3 m canopy height model (CHM) and laser-metric rasters calculated from all returns within the forest boundary. Plot-level covariates summarize these data over the sampled forest area of each field plot. For spatial prediction, we partition the forest into square cells with side length $\sqrt{\pi\,20^2}\approx35.45$ m, so a complete cell has the same area as a full field plot. Cells are clipped to the forest boundary, and the same covariates are calculated over each cell's forested area. The prediction grid comprises 4,812 cells with at least 95\% UAV coverage and spans 512.77 ha across all 249 stands. \figref{fig:weissekrug_data_map} shows the forest stands, field plots, and upper-canopy covariate.

We use three UAV covariates, denoted $x_{\mathrm{p95}}$, $x_{\mathrm{cover}}$, and $x_{\mathrm{mid}}$. All three weight raster cells by the area of each cell falling within the forest-clipped field plot or prediction cell. The first is the 95th percentile of CHM cell heights and represents upper-canopy development. The second summarizes the fraction of laser returns above 2 m and represents vegetation cover. The third summarizes the fraction of returns in the 10--20 m height band relative to all returns above 2 m, providing information about vegetation below the upper canopy in taller stands. The two return-based covariates use means of cell-level return fractions. Covariates are standardized to zero mean and unit variance using the field-plot means and standard deviations.

Exploratory data analysis guided the choice of this small covariate set, which represents complementary aspects of forest structure likely to inform tree density and species-by-DBH composition. Although covariate selection can affect predictive performance, our focus is on evaluating the modeling framework rather than identifying an optimal covariate set.

\subsection{Candidate models and prediction summaries}
\label{sec:candidate_model_ladder}
\label{sec:dbh_diversity}

We compare the eight models in \tabref{tab:model_ladder} using the first simulated population and the Weisse Krug data. These candidate models vary in their use of covariates, ordered-category smoothing, and spatial effects. All use the fixed-area negative-binomial likelihood and probit continuation-ratio composition model, with species as the leading margin and DBH class as the ordered conditional margin. Models M1--M5 estimate $r_M$, whereas M6--M8 fix $r_M=1000$ and include a spatial magnitude effect. Spatial components are fitted using the NNGP approximation in \secref{sec:computation}.

\begin{table}[!ht]
\centering
\scriptsize
\renewcommand{\arraystretch}{1.15}
\caption{Model specifications for the applications. The covariate set is $x_{\mathrm{p95}}$, $x_{\mathrm{cover}}$, and $x_{\mathrm{mid}}$; every regression includes an intercept. RW2 denotes second-order random-walk smoothing over ordered DBH classes.}
\label{tab:model_ladder}
\begin{tabularx}{\textwidth}{@{}c*{4}{>{\raggedright\arraybackslash}X}@{}}
\toprule
Model & Magnitude component & Species-group composition & Conditional DBH composition & Purpose \\
\midrule
M1 & Intercept only; $r_M$ estimated & Intercept only & Intercept only & Baseline negative-binomial model. \\
M2 & Same as M1 & Intercept only & RW2 intercepts & Adds ordered-DBH smoothing. \\
M3 & Covariates; $r_M$ estimated & Intercept only & Intercept only & Adds covariates for overall stem density. \\
M4 & Same as M3 & Intercept only & RW2 intercepts & Combines magnitude covariates and DBH smoothing. \\
M5 & Same as M3 & Covariates & Covariates; RW2 smoothing of intercepts and slopes & Adds covariates to both composition blocks. \\
M6 & Covariates and spatial magnitude NNGP; $r_M=1000$ & Covariates & Same as M5 & Adds spatial structure to overall stem density. \\
M7 & Same as M6 & Covariates and one spatial factor ($K=1$) & Same as M5 & Adds spatial structure to species composition. \\
M8 & Same as M6 & Covariates and two spatial factors ($K=2$) & Same as M5 & Tests a second species-composition factor. \\
\bottomrule
\end{tabularx}
\end{table}

M7 uses European beech as its loading anchor, and M8 uses European beech followed by Scots pine. These are the two most abundant species groups in the inventory (\tabref{tab:real_inventory_species}).

Forest and stand estimates are obtained by aggregating joint posterior predictive densities as described in \secref{sec:prediction_aggregation}. Predicted counts use the forested area of each prediction cell. Cell densities are weighted by their intersection areas with each stand, treating density as constant within each cell. Stand densities use the summed intersection areas as denominators. Forest summaries use the combined area of the prediction cells. Posterior means and pointwise 95\% predictive intervals summarize the aggregated densities and derived quantities.

We summarize structural diversity by applying Shannon entropy to DBH-class proportions \citep{staudhammer2001introduction} and expressing it as an effective number of equally common classes, the order-one Hill diversity \citep{hill1973diversity,jost2006entropy}. The DBH-class densities in \eqnref{eq:stand-derived-summaries} are obtained by summing over species. For draws with positive stand stem density, the predicted proportion in DBH class $b=1,\ldots,J_B$ for stand $S$ is
\begin{equation}
  \widetilde p_{S+b}^{\,(\ell)}
  =
  \frac{\widetilde d_{S+b}^{\,(\ell)}}
       {\widetilde d_S^{\,(\ell)}}.
  \label{eq:stand-dbh-proportion}
\end{equation}
The DBH-class diversity for draw $\ell$ is
\begin{equation}
  \widetilde D_S^{\,(\ell)}
  =
  \exp\left\{
    -\sum_{b=1}^{J_B}
    \widetilde p_{S+b}^{\,(\ell)}
    \log \widetilde p_{S+b}^{\,(\ell)}
  \right\},
  \label{eq:dbh-diversity}
\end{equation}
with zero-proportion DBH classes omitted from the sum. The metric ranges from 1 when all trees occupy a single DBH class to $J_B$ when all classes are equally represented. Larger values indicate that trees are spread across more classes or distributed more evenly among them, while smaller values indicate concentration in fewer classes.

\subsection{Simulation study}
\label{sec:simulation_study}

\subsubsection{Population generation and fitting}
\label{sec:repeated_simulation}

We generate 200 populations under M8, using the inventory's 151 plot locations and the 4,812 prediction cells across the 249 forest stands. This model-based assessment holds the locations, covariates, regression coefficients, factor loadings, and spatial covariance parameters fixed across populations.

Generating regression coefficients are posterior means from a regression-only fit to the Weisse Krug data, using the three covariates and RW2 smoothing of conditional DBH coefficients. The M8 generating parameter values are given in \tabref{tab:simulated_m8_scalar_parameter_recovery} and \figrefrange{fig:simulated_m8_species_coefficient_recovery}{fig:simulated_m8_dbh_mid_recovery} in Supplementary Section~\ref{sec:simulation_generating_parameters}.

To generate each population, we first draw one magnitude spatial effect and two unit-variance species-composition factors jointly over the plots and prediction grid from independent full exponential GPs. The magnitude regression and its spatial effect determine expected stem density $M_i$ through \eqnref{eq:magnitude-model}. The species-composition regression and spatial factors determine species probabilities, while the conditional DBH regressions determine DBH-class probabilities within species. Their product gives the species-by-DBH probabilities $p_{iab}$ in \eqnref{eq:two-way-factorization}. Multiplying these probabilities by $M_i$ gives expected species-by-DBH densities.

At each plot, the sampled area converts expected stem density to an expected tree count, $\plotarea_iM_i$. We draw the total count $N_i$ from the negative-binomial model in \eqnref{eq:total-count}, with $r_M=1000$, and allocate those trees among species-by-DBH categories using the multinomial model in \eqnref{eq:composition-count} with probabilities $p_{iab}$. We also generate one negative-binomial total and multinomial allocation at each prediction location using its forested cell area. Dividing these counts by cell area and aggregating with the same cell--stand intersection areas used for prediction defines the grid-based forest and stand densities used as simulation truth; the resulting stand tables also define true DBH-class diversity.

We fit M8 to the 151 plots in each population using the NNGP approximation and generate posterior predictive densities jointly over the population grid using \eqnref{eq:predictive-counts}. Prior and computational settings are given in Supplementary Section~\ref{app:application_computation}.

\subsubsection{Estimation performance}

We compare posterior means and pointwise 95\% predictive intervals with known population quantities across the 200 simulated populations. Bias, relative bias, root mean squared error (RMSE), coverage, and mean interval width summarize performance; definitions and Monte Carlo standard errors are given in Supplementary Section~\ref{sec:simulation_metrics}.

We assess five estimation targets: forest-wide densities for the 15 DBH classes using \eqnref{eq:stand-derived-summaries} and the 180 species-by-DBH cells using \eqnref{eq:stand-density}, these same two density summaries for each of the 249 stands, and stand DBH-class diversity using \eqnrefs{eq:stand-dbh-proportion}{eq:dbh-diversity}. \tabref{tab:simulated_repeated_performance} presents the results in this order. Across the five estimation targets, relative biases were small and predictive-interval coverage was near the nominal 95\% level.

\begin{table}[!ht]
\centering
\footnotesize
\setlength{\tabcolsep}{3pt}
\caption{M8 estimation performance across 200 simulated populations, with Monte Carlo standard errors in parentheses. Density errors and interval widths are in trees/ha; diversity errors and widths are in effective numbers of DBH classes. Relative bias and coverage are percentages, with standard errors in percentage points. Coverage concerns pointwise 95\% predictive intervals for the aggregated population quantities.}
\label{tab:simulated_repeated_performance}
\begin{tabular*}{\textwidth}{@{\extracolsep{\fill}}lrrrrr@{}}
\toprule
Target & \multicolumn{1}{c}{Bias} & \multicolumn{1}{c}{\shortstack{Relative\\bias (\%)}} & \multicolumn{1}{c}{RMSE} & \multicolumn{1}{c}{\shortstack{Coverage\\(\%)}} & \multicolumn{1}{c@{}}{\shortstack{Mean\\width}} \\
\midrule
\multicolumn{6}{@{}l}{\textbf{Forest}} \\
\quad DBH density & -0.413 (0.100) & -1.21 (0.30) & 3.169 (0.159) & 95.4 (0.5) & 8.575 (0.114) \\
\quad Species-by-DBH density & -0.034 (0.008) & -1.21 (0.30) & 0.722 (0.023) & 95.3 (0.2) & 1.464 (0.016) \\
\addlinespace
\multicolumn{6}{@{}l}{\textbf{Stand}} \\
\quad DBH density & -0.327 (0.124) & -0.87 (0.36) & 28.574 (0.813) & 95.0 (0.1) & 45.184 (0.585) \\
\quad Species-by-DBH density & -0.027 (0.010) & -0.87 (0.36) & 6.844 (0.212) & 97.0 (0.0) & 5.840 (0.068) \\
\quad DBH-class diversity & -0.077 (0.007) & -0.93 (0.08) & 0.505 (0.003) & 93.8 (0.2) & 1.784 (0.007) \\
\bottomrule
\end{tabular*}
\end{table}
\FloatBarrier

We also assess coverage of pointwise 95\% credible intervals for the generating parameters and realized spatial effects at the sampled plots. Coverage is near the nominal level for the regression coefficients and the magnitude spatial parameters (\tabref{tab:simulated_repeated_parameter_coverage}).

\begin{table}[!ht]
\centering
\small
\caption{M8 pointwise 95\% credible-interval coverage for parameters and spatial effects across 200 simulated populations. Numbers of values refer to each population. Monte Carlo standard errors in parentheses.}
\label{tab:simulated_repeated_parameter_coverage}
\begin{tabular}{@{}lrr@{}}
\toprule
Parameter or latent quantity & \multicolumn{1}{c}{\shortstack{Number of\\values}} & \multicolumn{1}{c@{}}{\shortstack{Coverage\\(\%)}} \\
\midrule
Magnitude coefficients & 4 & 96.4 (0.7) \\
Magnitude spatial variance & 1 & 95.5 (1.5) \\
Magnitude spatial decay & 1 & 94.0 (1.7) \\
Species coefficients & 44 & 93.4 (0.4) \\
Conditional DBH coefficients & 672 & 96.3 (0.1) \\
Factor loadings & 21 & 92.1 (0.6) \\
Composition-factor decays & 2 & 87.2 (1.7) \\
\addlinespace
Magnitude spatial effects & 151 & 93.9 (0.8) \\
Composition factor 1 scores & 151 & 87.5 (1.2) \\
Composition factor 2 scores & 151 & 93.4 (0.3) \\
\bottomrule
\end{tabular}
\end{table}
\FloatBarrier

A detailed multichain analysis of the first simulated population is presented in Supplementary Section~\ref{sec:simulation_generating_parameters}. It examines parameter estimates and chain convergence and illustrates forest density and stand diversity predictions. Cross-validation ranks the data-generating M8 model first by CRPS among the candidates, illustrating the usefulness of this criterion for guiding model choice.

\subsection{Weisse Krug analysis}
\label{sec:real_data_analysis}

\subsubsection{Model choice}

We next apply the cross-validation approach described in \secref{sec:model_choice} to the Weisse Krug data to select among the candidate models in \tabref{tab:model_ladder}. For each model, we predict species-by-DBH densities at each of the 151 plots using a fit that excludes that plot. \tabref{tab:real_cv_species_dbh} summarizes each candidate model's predictive performance for the full species-by-DBH table. We consider these results together with model complexity and MCMC diagnostics. Prior specifications, MCMC settings, and detailed diagnostics are given in Supplementary Section~\ref{app:application_computation}.

\begin{table}[!ht]
\centering
\small
\setlength{\tabcolsep}{3pt}
\caption{Leave-one-plot-out cross-validation of species-by-DBH density for the 151 Weisse Krug plots and 180 (12 species $\times$ 15 DBH classes) cells per plot. Mean error, RMSPE, CRPS, and mean interval width are in trees/ha; relative mean error and coverage are percentages. Coverage and width describe central 95\% posterior predictive intervals for held-out observations.}
\label{tab:real_cv_species_dbh}
\begin{tabular*}{\textwidth}{@{\extracolsep{\fill}}lrrrrrr@{}}
\toprule
Model & \multicolumn{1}{c}{\shortstack{Mean\\error}} & \multicolumn{1}{c}{\shortstack{Relative\\mean error (\%)}} & \multicolumn{1}{c}{RMSPE} & \multicolumn{1}{c}{CRPS} & \multicolumn{1}{c}{\shortstack{Coverage\\(\%)}} & \multicolumn{1}{c@{}}{\shortstack{Mean\\width}} \\
\midrule
M1 & 0.006 & 0.29 & 12.815 & 2.044 & 97.8 & 10.120 \\
M2 & 0.006 & 0.29 & 12.814 & 2.042 & 97.8 & 10.124 \\
M3 & 0.082 & 4.06 & 13.116 & 2.155 & 97.4 & 9.626 \\
M4 & 0.082 & 4.05 & 13.115 & 2.154 & 97.4 & 9.636 \\
M5 & 0.082 & 4.05 & 12.739 & 1.932 & 97.7 & 8.632 \\
M6 & 0.090 & 4.44 & 12.696 & 1.925 & 97.7 & 8.774 \\
M7 & 0.093 & 4.63 & 12.711 & 1.742 & 98.2 & 12.074 \\
M8 & 0.090 & 4.46 & 12.766 & 1.727 & 98.3 & 13.465 \\
\bottomrule
\end{tabular*}
\end{table}
\FloatBarrier

Adding covariates to the composition models reduces CRPS from 2.154 trees/ha for M4 to 1.932 for M5. The spatial magnitude effect in M6 gives a smaller further improvement, while the first spatial composition factor in M7 reduces CRPS to 1.742. M8 has the lowest CRPS, 1.727 trees/ha, but its second composition factor improves this score by only 0.9\% relative to M7.

Additional cross-validation comparisons for overall, species, and DBH-class densities are presented in Supplementary Section~\ref{sec:real_cv_targets}, together with observed-versus-predicted DBH-class densities. M7 and M8 also have the lowest CRPS for species and DBH-class densities, while M6--M8 perform similarly for overall density.

M7 and M8 have nearly identical RMSPE, and the intercept-only models M1 and M2 have the smallest relative mean errors. M8's multichain fit shows elevated $\widehat R$ values for many factor loadings, suggesting that its additional spatial structure is difficult to estimate from these data. We therefore select M7, which provides nearly the same CRPS with fewer parameters and better MCMC convergence (Supplementary Section~\ref{sec:real_parameter_estimates}). The following subsection presents the resulting estimates of forest species-by-DBH densities and stand DBH-class diversity.

\subsubsection{Forest and stand estimates}

\begin{figure}[!ht]
\centering
\includegraphics[width=\textwidth,height=0.70\textheight,keepaspectratio]{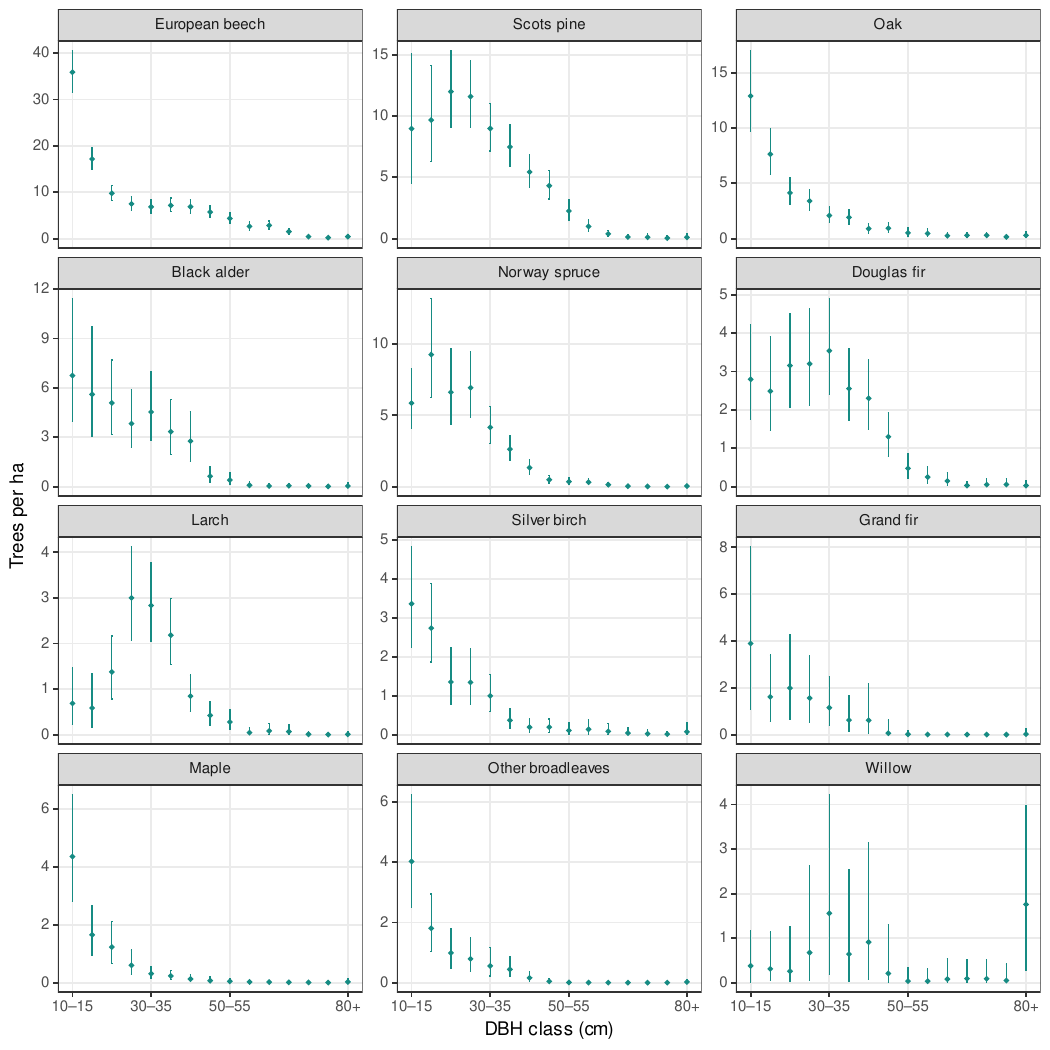}
\caption{M7 forest species-by-DBH density estimates for Weisse Krug. Points show posterior means and vertical bars show pointwise 95\% predictive intervals. Vertical scales vary by species; the terminal class includes all diameters at least 80 cm.}
\label{fig:real_forest_species_dbh_estimates}
\end{figure}

We fit M7 to all 151 plots. Parameter estimates are presented in \tabref{tab:real_scalar_parameters} and \figrefrange{fig:real_species_coefficients}{fig:real_dbh_mid}, with associated MCMC diagnostics in Supplementary Section~\ref{sec:real_parameter_estimates}. Joint prediction over the 4,812 prediction cells and subsequent aggregation give forest species-by-DBH densities and DBH-class diversity for the 249 stands. The estimated forest-wide stem density is 371.6 trees/ha, with a 95\% predictive interval of (346.0, 399.9). \figref{fig:real_forest_species_dbh_estimates} shows its allocation among species and DBH classes.

European beech and Scots pine contribute the largest estimated forest-wide densities, but their DBH distributions differ. Beech and oak densities are highest in the 10--15 cm class, whereas Scots pine and larch have their highest estimated densities in the 20--25 and 25--30 cm classes, respectively. Intervals are wide relative to estimated densities for grand fir and willow, consistent with their limited representation in the inventory: these species occur on only two and three plots, respectively (\tabref{tab:real_inventory_species}).

\begin{figure}[!ht]
\centering
\includegraphics[width=\textwidth]{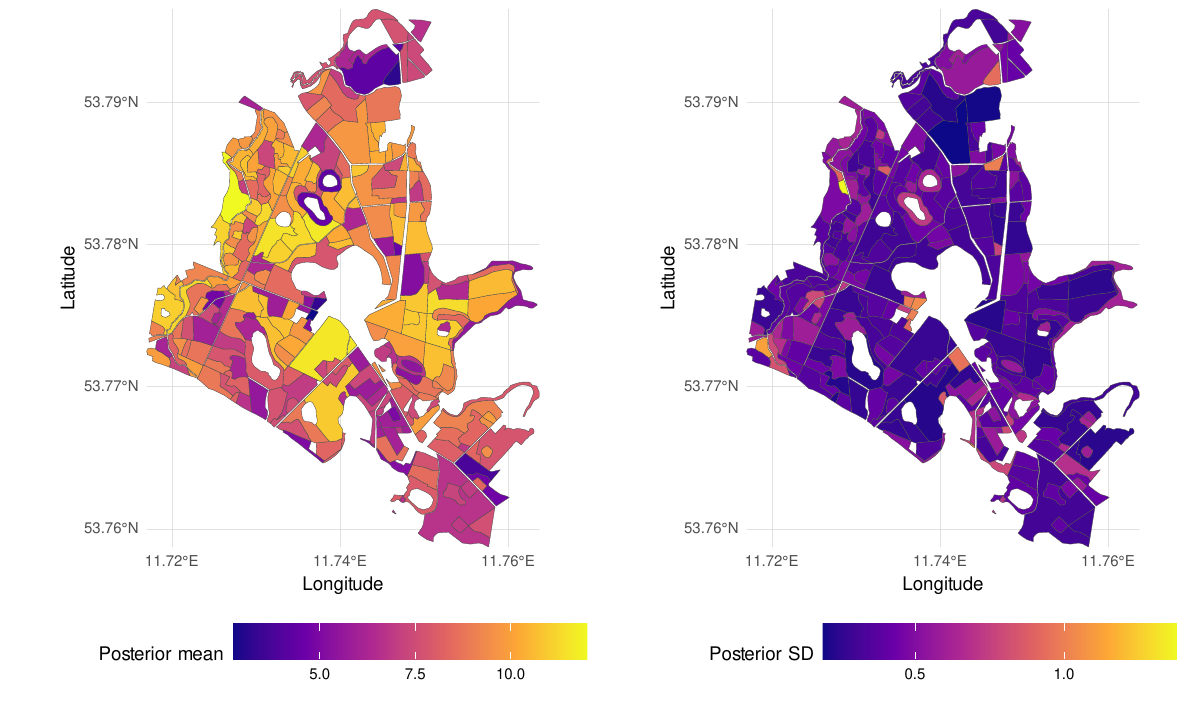}
\caption{M7 stand DBH-class diversity for Weisse Krug: posterior mean (left) and posterior standard deviation (right). Diversity is expressed as the effective number of equally common DBH classes.}
\label{fig:real_stand_diversity}
\end{figure}

Stand DBH-class diversity estimates range from 2.74 to 11.99 effective classes, with a median of 8.47 (\figref{fig:real_stand_diversity}). Lower values indicate that stems are concentrated in fewer DBH classes, while higher values indicate a broader and more even distribution across classes. Differences among nearby stands show local variation in size structure that is obscured by the forest-wide summaries. Posterior standard deviations range from 0.19 to 1.38, with a median of 0.42, and are generally small compared with the differences in estimated diversity across stands.

The model provides estimates for 139 stands without sampled plots, representing approximately 32\% of the forest area. Median posterior standard deviation of DBH-class diversity is 0.50 effective classes for these stands, compared with 0.34 for stands containing sampled plots. Together, the plot observations and LiDAR support species-by-DBH estimates for sampled and unsampled stands, extending the inventory's usefulness beyond forest-wide summaries. The accompanying uncertainty makes the lower precision in unsampled stands explicit.

\FloatBarrier
\section{Discussion and future work}\label{sec:discussion}

\subsection{Findings from the analyses}\label{sec:discussion_findings}

The magnitude-composition formulation connects estimation of overall abundance with estimation of species and size distributions, using covariates to inform both components. Building on the spatial stand-table approach of \citet{finley2014dynamic}, it gives abundance and allocation separate regression and random-effect structures while keeping category densities and their marginal sums consistent. Smoothing across ordered DBH coefficients and shared composition factors provide ways to estimate many category-specific quantities from a limited set of plots. These features support a flexible description of forest structure without requiring a particular parametric DBH distribution.

The repeated-population study shows small biases and predictive-interval coverage near nominal levels for forest and stand densities and the nonlinear summary of DBH-class diversity (\tabref{tab:simulated_repeated_performance}). These results, obtained using NNGP fitting and prediction for populations generated from full GPs, support the approximation under the simulated conditions, consistent with earlier findings on NNGP predictive performance \citep{datta2016hierarchical,Finley2019nngpcomp}.

The Weisse Krug analysis demonstrates the value of unit-level SAE for describing stand structure when field sampling is sparse. LiDAR covariates and spatial dependence allow observations from sampled plots to inform species-by-DBH estimates elsewhere in the forest. The resulting stand tables and diversity summaries reveal local variation in size structure that is obscured by forest-wide summaries. Reporting uncertainty alongside these estimates makes their differing precision explicit, particularly for stands without sampled plots.

The model comparisons also show the value of considering complementary predictive metrics alongside model complexity and convergence. Cross-validation favors the data-generating two-factor model in the first simulated population (\tabref{tab:simulated_cv_species_dbh}), whereas the second factor offers little predictive gain for Weisse Krug and is harder to estimate (\tabref{tab:real_cv_species_dbh}). The selected one-factor model retains some mixing limitations for grand-fir coefficients. Grand fir occurs on only two plots with similar covariates; additional trees at those locations do not provide the covariate variation needed to distinguish its regression effects (\tabref{tab:real_inventory_species}). Thus, information for species-specific regression depends on the number and covariate range of occupied plots as well as tree abundance.

\subsection{Spatiotemporal extensions}\label{sec:spatiotemporal_extensions}

Forest inventories covering large spatial extents, including national forest inventories, often distribute sampling effort across years using panels of permanent plots. Each panel contains a subset of plots scheduled for measurement in the same year, with a complete inventory cycle spanning all panels. The U.S. Department of Agriculture Forest Service's Forest Inventory and Analysis (FIA) program uses such a design, with measurement cycles generally spanning five to ten years depending on the region \citep{bechtold2005enhanced}. Combining these observations to estimate forest conditions for a specified year, or changes between years, motivates models that account for measurement year.

An extension of the magnitude-composition model would index observations and covariates by location and time and replace the spatial effects with spatiotemporal processes $v(\bs,t)$ and $\bw(\bs,t)$, where $t$ denotes time. These processes could use nonseparable space--time covariance functions such as those of \citet{gneiting2002nonseparable}, or covariance structures tailored to the multiple scales of dependence in large forest inventories \citep[see, e.g.,][]{may2025spatiotemporal}.

Joint prediction would then provide stand-tables, diversity estimates, and changes between years with associated uncertainty. Remotely sensed covariates acquired on multiple dates could provide additional information about these changes. Posterior stand-table draws could also inform growth-and-yield models that describe growth, mortality, and recruitment, such as the framework of \citet{itter2025growth}, carrying uncertainty in estimated forest structure into projections under alternative management scenarios.

\subsection{Covariates and sampling designs}\label{sec:covariates_sampling_designs}

We used the same three LiDAR covariates for magnitude, species allocation, and DBH allocation within species. The model permits different covariates for each component, allowing variable selection tailored to overall stem density, species discrimination, and size structure. A promising avenue for future work is to inform the magnitude component using remotely sensed estimates of stems per unit area from individual-tree detection with corrections for missed and false detections \citep{kostensalo2023recreating,kostensalo2026large}. LiDAR measurements of canopy structure may be particularly informative for DBH allocation, while hyperspectral data could supply complementary information for distinguishing species \citep{dalponte2012species}.

The framework also extends to horizontal point sampling, another commonly used forest inventory method, with magnitude representing basal-area density. The formulation and its prediction and aggregation steps are given in Supplementary Section~\ref{sec:HPS}. Further development should account for boundary effects on tree inclusion and identify appropriate support for auxiliary covariates \citep{kirchhoefer2017considerations}.

\section*{Significance statement}

Sparse forest inventories often cannot provide the species and tree-size information needed for decisions about individual stands. We develop a general magnitude-composition framework that combines field measurements with auxiliary information to estimate overall abundance and its distribution among categories. In forest inventory, this yields species-by-size stand-tables for sampled and unsampled stands, together with uncertainty in their entries and derived measures such as structural diversity. These summaries support silvicultural planning and assessment of forest structure at scales finer than the inventory alone can reliably describe, while making the precision of local estimates explicit.

\begin{acks}
Finley and Itter were supported by the USDA Forest Service Forest Inventory and Analysis Program, the Rocky Mountain Research Station, and the National Council for Air and Stream Improvement (NCASI) Partnership for Small Area Estimation. Spitta was supported by the State Forestry Agency of Mecklenburg--West Pomerania (Landesforstanstalt Mecklenburg-Vorpommern).

The authors thank Kai J\"utte and OGF GmbH for acquiring and providing the UAV-borne laser scanning data used in this study on behalf of the State Forestry Agency of Mecklenburg--West Pomerania.
\end{acks}

\section*{Code and data}
Prepared data and scripts reproduce the simulation study, cross-validation comparisons, and Weisse Krug analysis. The README provides software requirements and execution instructions.

\bibliographystyle{spr-ims-nameyear}
\bibliography{refs_AOAS}

\clearpage
\section*{Supplementary material}

\appendix
\setcounter{equation}{0}
\setcounter{table}{0}
\setcounter{figure}{0}
\renewcommand{\theHequation}{supp.\arabic{equation}}
\renewcommand{\theHtable}{supp.\arabic{table}}
\renewcommand{\theHfigure}{supp.\arabic{figure}}
\renewcommand{\theequation}{S\arabic{equation}}
\renewcommand{\thetable}{S\arabic{table}}
\renewcommand{\thefigure}{S\arabic{figure}}

\section{Posterior sampling and computation}
\label{app:sampler}

This section gives the sampler for the fixed-area magnitude model and the leading and conditional probit allocation models in \secref{sec:model} of the main text. For Gibbs updates, $\mid\cdot$ denotes conditioning on the data and the latest values of all other parameters and latent variables. Each additional process term has its own parameters and is updated in turn, holding the other terms fixed. Additional implementation details are provided in the technical documentation accompanying the \textsf{magComp} package \citep{finley2026magcomp}.

\subsection{Augmented likelihoods}
\label{sec:sampler_augmentation}

For magnitude, write $\eta_i=\log M_i$, $o_i=\log\plotarea_i-\log r_M$, and $\psi_i^{(M)}=\eta_i+o_i$. P\'olya--Gamma augmentation gives
\begin{equation}
 \omega_i^{(M)}\given\cdot\sim\PG(N_i+r_M,\psi_i^{(M)}),
 \qquad \kappa_i=(N_i-r_M)/2,
 \label{eq:supp_pg}
\end{equation}
with augmented likelihood kernel $\exp\{\kappa_i\psi_i^{(M)}-\omega_i^{(M)}(\psi_i^{(M)})^2/2\}$ \citep{polson2013polya}. Let $\bOmega_M=\diag(\omega_i^{(M)})$, $\bkappa=(\kappa_i)_i$, and $\bo=(o_i)_i$.

Within either composition block, let $j$ index its continuation-ratio rows. These are the $J_A-1$ leading rows or the $J_A(J_B-1)$ parent--child rows. Thus $y_{ij}$, $R_{ij}$, and $\psi_{ij}$ below denote the corresponding count, remaining count, and linear predictor for plot $i$ and row $j$; in the conditional block, $j$ specifies both parent and child categories. For each of the $R_{ij}$ trees available at this allocation step, indexed by $t=1,\ldots,R_{ij}$, draw independent latent normal variables
\begin{equation}
 \zeta_{ijt}\given\cdot\sim
 \begin{cases}
  \N(\psi_{ij},1)\text{ restricted to }(0,\infty),&t\leq y_{ij},\\
  \N(\psi_{ij},1)\text{ restricted to }(-\infty,0],&y_{ij}<t\leq R_{ij},
 \end{cases}
 \label{eq:supp_probit}
\end{equation}
and retain their sum $S_{ij}=\sum_{t=1}^{R_{ij}}\zeta_{ijt}$. The augmented kernel is $\exp\{S_{ij}\psi_{ij}-R_{ij}\psi_{ij}^2/2\}$ \citep{albert1995bayesian,rouder2005introduction}.

\subsection{Normal draws using Cholesky solves}
\label{sec:sampler_solves}

Every unrestricted normal update below specifies a precision $\bQ$ and linear term $\bb$, defining the full conditional
\begin{equation}
 p(\boldsymbol\vartheta\given\cdot)\propto
 \exp\{-\tfrac12\boldsymbol\vartheta\tr\bQ\boldsymbol\vartheta
                  +\bb\tr\boldsymbol\vartheta\}.
 \label{eq:supp_normal_kernel}
\end{equation}
Its mean $\bmM$ solves $\bQ\bmM=\bb$. We use the following draw operations, which form no posterior covariance inverse. For a dense precision, $\mathcal G_D(\bQ,\bb)$ factors $\bQ=\bL\bL\tr$ and computes
\begin{equation}
 \begin{gathered}
  \bL\ba=\bb,\qquad \bL\tr\bmM=\ba,\qquad
  \bL\tr\be=\bepsilon,\qquad \bepsilon\sim\N(\bzero,\bI),\\
  \boldsymbol\vartheta=\bmM+\be.
 \end{gathered}
 \label{eq:supp_dense_draw}
\end{equation}
LAPACK Cholesky factorization and solves compute the mean; a BLAS triangular solve supplies the random component \citep{anderson1999lapack,blackford2002blas}.

For a sparse precision, $\mathcal G_S(\bQ,\bb)$ uses CHOLMOD's permutation $\boldsymbol\Pi$ and factorization $\boldsymbol\Pi\bQ\boldsymbol\Pi\tr=\bL\bL\tr$. The corresponding computation is
\begin{equation}
 \begin{gathered}
  \bL\ba=\boldsymbol\Pi\bb,\qquad \bL\tr\bmM_{\mathrm{perm}}=\ba,\qquad
  \bL\tr\be=\bepsilon,\qquad \bepsilon\sim\N(\bzero,\bI),\\
  \boldsymbol\vartheta=\boldsymbol\Pi\tr(\bmM_{\mathrm{perm}}+\be).
 \end{gathered}
 \label{eq:supp_sparse_draw}
\end{equation}
CHOLMOD reuses symbolic analysis of the permutation and factor structure for a fixed sparsity pattern, and recomputes the numerical factorization for each current precision \citep{chen2008cholmod}. This permutation is distinct from the spatial ordering used to define neighbors.

\subsection{NNGP prior precision}
\label{sec:sampler_nngp}

For each spatial term, order plots as described in \secref{sec:computation} of the main text and let $\mathcal N_i$ contain up to $m$ nearest preceding locations. In this subsection vectors are in that order. For unit-scale exponential correlation $\bR_\phi$, compute neighbor weights and conditional variances by solving
\begin{equation}
 \bR_{\mathcal N_i,\mathcal N_i}\bb_i=\bR_{\mathcal N_i,i},
 \qquad F_i=1-\bR_{i,\mathcal N_i}\bb_i.
 \label{eq:supp_neighbors}
\end{equation}
The solves use Cholesky factors of matrices of size at most $m\times m$. Let $\bB_\phi$ place $\bb_i\tr$ in row $i$ at its neighbor columns, with zeros elsewhere; $F_1=1$. The NNGP precision and quadratic form are
\begin{equation}
 \begin{aligned}
  \bH_\phi&=(\bI-\bB_\phi)\tr\diag(F_i^{-1})(\bI-\bB_\phi),\\
  T_\phi(\bv)&=\sum_i
    \frac{(v_i-\bb_i\tr\bv_{\mathcal N_i})^2}{F_i}
    =\bv\tr\bH_\phi\bv.
 \end{aligned}
 \label{eq:supp_nngp_precision}
\end{equation}
For a process with variance $s^2$, the precision is $s^{-2}\bH_\phi$ and its log density, up to a constant, is
\begin{equation}
 G(\bv;\phi,s^2)=-\tfrac12\left\{n\log s^2+
              \sum_i\log F_i+s^{-2}T_\phi(\bv)\right\}.
 \label{eq:supp_spatial_target}
\end{equation}
Use $s^2=\sigma_v^2$ for magnitude and $s^2=1$ for each composition factor. These constructions follow the NNGP representation of \citet{datta2016hierarchical,Finley2019nngpcomp}.

Neighbor identities are computed once, and the sparse precision is assembled directly from the conditional representation. Neighbor weights and variances are recomputed for decay proposals. The log determinant and quadratic form in \eqnref{eq:supp_spatial_target} use the conditional representation without a large matrix factorization. Latent-vector draws use \eqnref{eq:supp_sparse_draw}; their factorization cost depends on sparse fill-in.

\subsection{One sampler sweep}
\label{sec:sampler_sweep}

The following steps give one iteration in implementation order. An offset excludes only the quantity currently being updated and otherwise uses the latest parameter values. Write $c_i$ for a process term's covariate multiplier, equal to one for an intercept term. Suppress process-term and allocation-block superscripts where the same update applies to each. Independent standard normal draws below are denoted by $\epsilon$.

\begin{enumerate}
\item \textbf{Negative-binomial size, when estimated.}
Update $r_M$ using the marginal likelihood and log-scale Metropolis step in \secref{sec:sampler_metropolis}. Immediately refresh the magnitude augmentation in Step~2 after an estimated-size update.

\item \textbf{Magnitude augmentation and regression.}
Draw \eqnref{eq:supp_pg} using the current linear predictors and size. With $\bZ$ having rows $\bz_i\tr$ and $\boldf=(f_i)_i$ collecting the sums of magnitude random-effect contributions, set
\begin{equation}
 \begin{aligned}
  \bQ_\beta&=\sigma_\beta^{-2}\bI+\bZ\tr\bOmega_M\bZ,\\
  \bb_\beta&=\bZ\tr\{\bkappa-\bOmega_M(\bo+\boldf)\},
  \qquad \bbeta\leftarrow\mathcal G_D(\bQ_\beta,\bb_\beta).
 \end{aligned}
 \label{eq:supp_beta}
\end{equation}
Keep these augmentation weights through all magnitude-process updates.

\item \textbf{Magnitude process terms.}
Visit each included term in model order. Let $o_i^{(-)}$ be its excluded-term offset in $\psi_i^{(M)}$, including $\log\plotarea_i-\log r_M$. For a group term, update each $\delta_g$ and then its variance using
\begin{equation}
 \begin{aligned}
  Q_g&=\sigma_\delta^{-2}+\sum_{i:g(i)=g}\omega_i^{(M)}c_i^2,\qquad
  b_g=\sum_{i:g(i)=g}c_i\{\kappa_i-\omega_i^{(M)}o_i^{(-)}\},\\
  \delta_g&\leftarrow b_g/Q_g+\epsilon/\sqrt{Q_g},\qquad
  \sigma_\delta^2\given\cdot\sim
   \IG\left(a_\delta+n_\delta/2,b_\delta+\tfrac12\sum_g\delta_g^2\right),
 \end{aligned}
 \label{eq:supp_mag_iid}
\end{equation}
where $n_\delta$ is the number of groups for this term. For a spatial term, jointly draw its entire location vector and then its variance,
\begin{equation}
 \begin{aligned}
  \bQ_v&=\sigma_v^{-2}\bH_{\phi_v}+\diag(\omega_i^{(M)}c_i^2),\qquad
  b_{vi}=c_i\{\kappa_i-\omega_i^{(M)}o_i^{(-)}\},\\
  \bv&\leftarrow\mathcal G_S(\bQ_v,\bb_v),\qquad
  \sigma_v^2\given\cdot\sim
    \IG\left(a_\sigma+n/2,b_\sigma+\tfrac12T_{\phi_v}(\bv)\right).
 \end{aligned}
 \label{eq:supp_mag_spatial}
\end{equation}
Then update its estimated decay using \secref{sec:sampler_metropolis}. Inverse-gamma distributions use shape and scale as in the main text.

\item \textbf{Composition blocks.}
For each block, leading before conditional, perform Steps~4(a)--4(c) before moving to the next block.
\begin{enumerate}
\item Draw its latent normals and sums from \eqnref{eq:supp_probit}. Keep these sums through the block's coefficient and factor updates.

\item Update the intercept vector and then each slope vector. Let $X_{iq}$ denote its design entry, with $X_{i0}=1$, and let $\btheta_q$ collect the corresponding coefficients over all rows $j$, with $\theta_{0j}=\alpha_{0j}$ and $\theta_{qj}$ the $q$th slope in $\balpha_j$ for $q>0$. Its prior precision $\bP_q$ is the independent-prior precision or the smoothing precision in \eqnref{eq:smoothing-prior} of the main text. In conditional blocks this precision is block diagonal over parents. For $o_{ij}^{(-q)}=\psi_{ij}-X_{iq}\theta_{qj}$, form
\begin{equation}
 \begin{aligned}
  D_{qj}&=\sum_i R_{ij}X_{iq}^2,\qquad
  b_{qj}=\sum_i X_{iq}\{S_{ij}-R_{ij}o_{ij}^{(-q)}\},\\
  \bQ_q&=\bP_q+\diag(D_{qj}),\qquad
  \btheta_q\leftarrow\mathcal G_D(\bQ_q,\bb_q).
 \end{aligned}
 \label{eq:supp_alpha}
\end{equation}
The draw includes all parent-specific coefficient rows together, using a dense factorization. Smoothing connects ordered rows within each parent, without a penalty across parents.

\item For each factor term, update its loadings followed by its scores and estimated decays as follows.

\emph{Loadings.} Visit estimated loadings by row and then by factor column. For a spatial loading $\lambda_{jk}$, write $\xi_i=c_iw_k(\bs_i)$ and $o_{ij}^{(-jk)}=\psi_{ij}-\xi_i\lambda_{jk}$. Its scalar conditional precision and mean are
\begin{equation}
 Q_{jk}=c_\lambda^{-1}+\sum_iR_{ij}\xi_i^2,\qquad
 m_{jk}=\frac{\sum_i \xi_i\{S_{ij}-R_{ij}o_{ij}^{(-jk)}\}}{Q_{jk}}.
 \label{eq:supp_loadings}
\end{equation}
Draw $\lambda_{jk}\leftarrow m_{jk}+\epsilon/\sqrt{Q_{jk}}$ for unrestricted entries. At an anchor diagonal, draw from $\N(m_{jk},Q_{jk}^{-1})$ restricted to $(0,\infty)$. For an unstructured loading, use $\xi_i=c_iu_{h(i),k}$ and replace $\lambda$ and $c_\lambda$ with $\gamma$ and $c_\gamma$.

\emph{Unstructured scores.} Let $\boldsymbol\xi_{ij}=c_i\bgamma_j$, and let $o_{ij}^{(-\mathrm{term})}$ exclude this entire factor term. Jointly draw the $K$ scores within each group $h$ using
\begin{equation}
 \begin{aligned}
  \bQ_h&=\bI_K+\sum_{i:h(i)=h}\sum_jR_{ij}\boldsymbol\xi_{ij}\boldsymbol\xi_{ij}\tr,\\
  \bb_h&=\sum_{i:h(i)=h}\sum_j\boldsymbol\xi_{ij}
                     \{S_{ij}-R_{ij}o_{ij}^{(-\mathrm{term})}\},\qquad
  \bu_h\leftarrow\mathcal G_D(\bQ_h,\bb_h).
 \end{aligned}
 \label{eq:supp_iid_factors}
\end{equation}

\emph{Spatial scores.} Update each full spatial column $\bw_k=(w_k(\bs_i))_i$ in turn, conditional on the other columns. Let $\xi_{ij}=c_i\lambda_{jk}$ and $o_{ij}^{(-k)}=\psi_{ij}-\xi_{ij}w_k(\bs_i)$. Set
\begin{equation}
 \begin{aligned}
  D_{ik}&=\sum_jR_{ij}\xi_{ij}^2,\qquad
  b_{ik}=\sum_j\xi_{ij}\{S_{ij}-R_{ij}o_{ij}^{(-k)}\},\\
  \bQ_k&=\bH_{\phi_{w,k}}+\diag(D_{ik}),\qquad
  \bw_k\leftarrow\mathcal G_S(\bQ_k,\bb_k).
 \end{aligned}
 \label{eq:supp_spatial_factors}
\end{equation}
After all columns, update each estimated factor decay as below.
\end{enumerate}
\end{enumerate}

All composition sums use the rows of the current block. In particular, the score updates in \eqnrefs{eq:supp_iid_factors}{eq:supp_spatial_factors} sum over all parent--child rows for a conditional block. Retain the parameter state after the complete sweep, according to the burn-in and thinning settings in \secref{sec:application_mcmc}.

\subsection{Metropolis updates and numerical details}
\label{sec:sampler_metropolis}

Under the uniform decay prior, the conditional density is proportional to $\exp\{G(\bv;\phi,s^2)\}$ for $\phi_L<\phi<\phi_U$. Use a normal random walk on the bounded-logit scale,
\[
 \begin{aligned}
  \eta_\phi&=\log\frac{\phi-\phi_L}{\phi_U-\phi},\qquad
  \eta_\phi^\star=\eta_\phi+s_\phi\epsilon,\\
  \phi^\star&=\phi_L+(\phi_U-\phi_L)\operatorname{logit}^{-1}(\eta_\phi^\star),
 \end{aligned}
\]
and accept with probability
\begin{equation}
 \min\left[1,\exp\left\{
 G(\bv;\phi^\star,s^2)-G(\bv;\phi,s^2)
 +\log\frac{(\phi^\star-\phi_L)(\phi_U-\phi^\star)}
                {(\phi-\phi_L)(\phi_U-\phi)}\right\}\right].
 \label{eq:supp_decay_mh}
\end{equation}
Use the updated magnitude vector and variance, or the updated factor column with $s^2=1$. The final term is the log ratio of the inverse transformation's Jacobians.

When size is estimated, write its log likelihood up to terms independent of $r$ and the log-scale target as
\begin{equation}
 \begin{aligned}
 \log\mathcal L_M(r)&=\sum_i\left\{\log\Gamma(N_i+r)-\log\Gamma(r)
       +r\log\frac{r}{r+\plotarea_iM_i}+N_i\log\frac{\plotarea_iM_i}{r+\plotarea_iM_i}\right\},\\
 H(r)&=\log\mathcal L_M(r)+(a_r-1)\log r-b_r r+\log r.
 \end{aligned}
 \label{eq:supp_size_target}
\end{equation}
Propose $\log r_M^\star=\log r_M+s_r\epsilon$ and accept with probability $\min[1,\exp\{H(r_M^\star)-H(r_M)\}]$. This update uses $p(r_M\mid\bN,\bM)$ with the previous P\'olya--Gamma variables integrated out. Refreshing them immediately afterward is part of the sampler. The final $\log r$ in $H$ corrects for the log-scale proposal.

Metropolis proposals are tuned adaptively. Analysis-specific chain lengths are given in \secref{sec:application_mcmc}.

Magnitude augmentation uses the hybrid P\'olya--Gamma generator in the \textsf{BayesLogit} R package \citep{polson2026bayeslogit}, which includes numerical approximations to the target distribution in \eqnref{eq:supp_pg}.

\section{Supplementary assessment metric definitions}
\label{app:assessment_metrics}

\subsection{Cross-validation}
\label{sec:cv_metrics}

For each candidate model, let $O_e$, $e=1,\ldots,L$, be the observed species-by-DBH densities at held-out plots, calculated as cell counts divided by sampled plot area. The comparisons in Tables~\ref{tab:real_cv_species_dbh} and~\ref{tab:simulated_cv_species_dbh} include $L=151\times180$ plot-by-species-by-DBH entries.

Let $\widehat O_e$ be the posterior predictive mean density from the fit excluding that entry's plot. With error $d_e=\widehat O_e-O_e$, mean error (ME), relative mean error (RME), and root mean squared prediction error (RMSPE) are
\begin{align}
  \mathrm{ME}
  &=\frac{1}{L}\sum_{e=1}^{L}d_e,
  \label{eq:cv-mean-error}
  \\
  \mathrm{RME}
  &=100\frac{\sum_{e=1}^{L}d_e}{\sum_{e=1}^{L}O_e},
  \label{eq:cv-relative-mean-error}
  \\
  \mathrm{RMSPE}
  &=\left\{\frac{1}{L}\sum_{e=1}^{L}d_e^2\right\}^{1/2}.
  \label{eq:cv-rmspe}
\end{align}

We calculate CRPS from posterior predictive draws for each entry and report its average over the $L$ entries, using the definition in \citet[Section~4.2]{gneiting2007strictly} with lower scores indicating better prediction.

Let $Q_e(0.025)$ and $Q_e(0.975)$ be the posterior predictive quantiles defining the central 95\% interval. Coverage and mean interval width are
\begin{align}
  \mathrm{Cover}
  &=\frac{100}{L}\sum_{e=1}^{L}
    \bone\{Q_e(0.025)\leq O_e\leq Q_e(0.975)\},
  \label{eq:cv-coverage}
  \\
  \mathrm{Width}
  &=\frac{1}{L}\sum_{e=1}^{L}
    \{Q_e(0.975)-Q_e(0.025)\}.
  \label{eq:cv-width}
\end{align}
\subsection{Simulation assessment}
\label{sec:simulation_metrics}

For the single-population results in \secref{sec:simulation_generating_parameters}, we apply the error and interval formulas in \secref{sec:cv_metrics} to known population values, denoting root mean squared error by RMSE. Estimates and interval endpoints are posterior means and quantiles of aggregated posterior predictive density or diversity draws.

For population $r=1,\ldots,R$, let $T_{re}$ and $\widehat T_{re}$ be the true and estimated values for entries $e=1,\ldots,L_r$. Entries correspond to the density or diversity targets in each row of \tabref{tab:simulated_repeated_performance}. We first calculate ME, RME, coverage, and width within each population using $O_e=T_{re}$ and $\widehat O_e=\widehat T_{re}$, then average each metric across the $R=200$ populations. The averages of ME and RME estimate bias and relative bias, respectively. Repeated-population RMSE is
\begin{align}
  \mathrm{MSE}_r
  &=\frac{1}{L_r}\sum_{e=1}^{L_r}
    (\widehat T_{re}-T_{re})^2,
  \label{eq:simulation-mse}
  \\
  \mathrm{RMSE}_{\mathrm{sim}}
  &=\left\{\frac{1}{R}\sum_{r=1}^{R}\mathrm{MSE}_r\right\}^{1/2}.
  \label{eq:simulation-rmse}
\end{align}

For the parameter and spatial-effect coverage in \tabref{tab:simulated_repeated_parameter_coverage}, we apply \eqnref{eq:cv-coverage} to the corresponding credible intervals, averaging within each block and then across populations.

Monte Carlo standard errors use populations as independent repetitions. For a metric averaged across populations, $\bar q=R^{-1}\sum_r q_r$, let $s_q$ be the sample standard deviation of its within-population values. Let $s_{\mathrm{MSE}}$ be the sample standard deviation of the $\mathrm{MSE}_r$ values. The reported standard errors are
\begin{equation}
  \mathrm{MCSE}(\bar q)=\frac{s_q}{\sqrt R},
  \qquad
  \mathrm{MCSE}(\mathrm{RMSE}_{\mathrm{sim}})
  \simeq\frac{s_{\mathrm{MSE}}}{2\sqrt R\,\mathrm{RMSE}_{\mathrm{sim}}},
  \label{eq:simulation-mcse}
\end{equation}
where the RMSE expression uses the delta method.

\section{Supplementary simulation analysis}
\label{app:simulation_results}

\subsection{Detailed analysis of the first simulated population}
\label{sec:simulation_generating_parameters}

We examine the first of the 200 populations generated as described in \secref{sec:repeated_simulation} of the main text in more detail to illustrate model comparison and estimation of parameters, forest densities, and stand diversity for a single realization.

We begin by applying the cross-validation approach in \secref{sec:model_choice} of the main text to the candidates in \tabref{tab:model_ladder}. For each candidate, we predict species-by-DBH densities at each of the 151 plots with that plot excluded from fitting and covariates standardized using the remaining plots. \tabref{tab:simulated_cv_species_dbh} summarizes predictive performance, with CRPS as the primary criterion. Prior and MCMC settings are given in \appref{app:application_computation}.

\begin{table}[!ht]
\centering
\small
\setlength{\tabcolsep}{3pt}
\caption{Leave-one-plot-out cross-validation of species-by-DBH density for the 151 plots and 180 cells per plot in the first simulated population. Mean error, RMSPE, CRPS, and mean interval width are in trees/ha; relative mean error and coverage are percentages. Coverage and width describe central 95\% posterior predictive intervals for held-out observations. Metric definitions are in \secref{sec:cv_metrics}.}
\label{tab:simulated_cv_species_dbh}
\begin{tabular*}{\textwidth}{@{\extracolsep{\fill}}lrrrrrr@{}}
\toprule
Model & \multicolumn{1}{c}{\shortstack{Mean\\error}} & \multicolumn{1}{c}{\shortstack{Relative\\mean error (\%)}} & \multicolumn{1}{c}{RMSPE} & \multicolumn{1}{c}{CRPS} & \multicolumn{1}{c}{\shortstack{Coverage\\(\%)}} & \multicolumn{1}{c@{}}{\shortstack{Mean\\width}} \\
\midrule
M1 & 0.006 & 0.21 & 11.754 & 2.378 & 97.9 & 13.538 \\
M2 & 0.006 & 0.22 & 11.752 & 2.380 & 97.9 & 13.606 \\
M3 & 0.103 & 3.59 & 11.651 & 2.383 & 97.7 & 13.322 \\
M4 & 0.102 & 3.56 & 11.647 & 2.382 & 97.8 & 13.362 \\
M5 & 0.102 & 3.57 & 11.092 & 1.847 & 98.8 & 12.035 \\
M6 & 0.053 & 1.86 & 8.697 & 1.675 & 98.9 & 10.848 \\
M7 & 0.054 & 1.89 & 8.232 & 1.591 & 99.2 & 12.059 \\
M8 & 0.054 & 1.89 & 8.250 & 1.582 & 99.2 & 12.304 \\
\bottomrule
\end{tabular*}
\end{table}
\FloatBarrier

Adding covariates to the composition models reduces CRPS from 2.382 trees/ha for M4 to 1.847 for M5. Including a spatial magnitude effect reduces it further to 1.675 for M6, and the first spatial composition factor reduces it to 1.591 for M7. M8, the generating model, has the lowest CRPS (1.582 trees/ha), an improvement of 0.5\% over M7. These results support including spatial composition factors but provide limited separation between one and two factors.

We examine M8 in detail below to assess estimation of its known generating parameters and population quantities. We fit M8 to all 151 plots using three chains of 25,000 iterations, discard the first 10,000 iterations of each chain, and retain every tenth subsequent draw. The resulting 4,500 draws are used for parameter summaries and joint prediction over the population grid.

\begin{figure}[p]
\centering
\includegraphics[width=\textwidth]{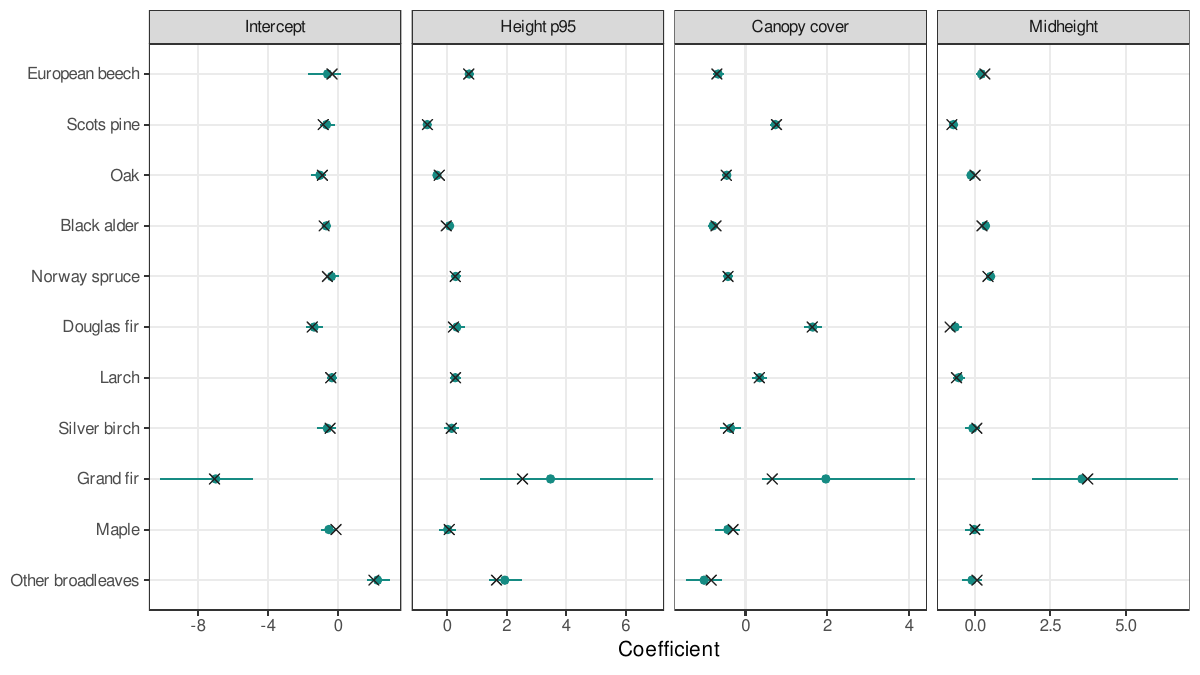}
\caption{Species-composition regression coefficient estimates for M8. Points and horizontal bars show posterior means and pointwise 95\% credible intervals; crosses mark generating values. Panels separate intercepts and the three standardized covariate coefficients, with horizontal scales varying by panel. The terminal species group has no coefficient row.}
\label{fig:simulated_m8_species_coefficient_recovery}
\end{figure}

\begin{figure}[p]
\centering
\includegraphics[width=\textwidth,height=0.73\textheight,keepaspectratio]{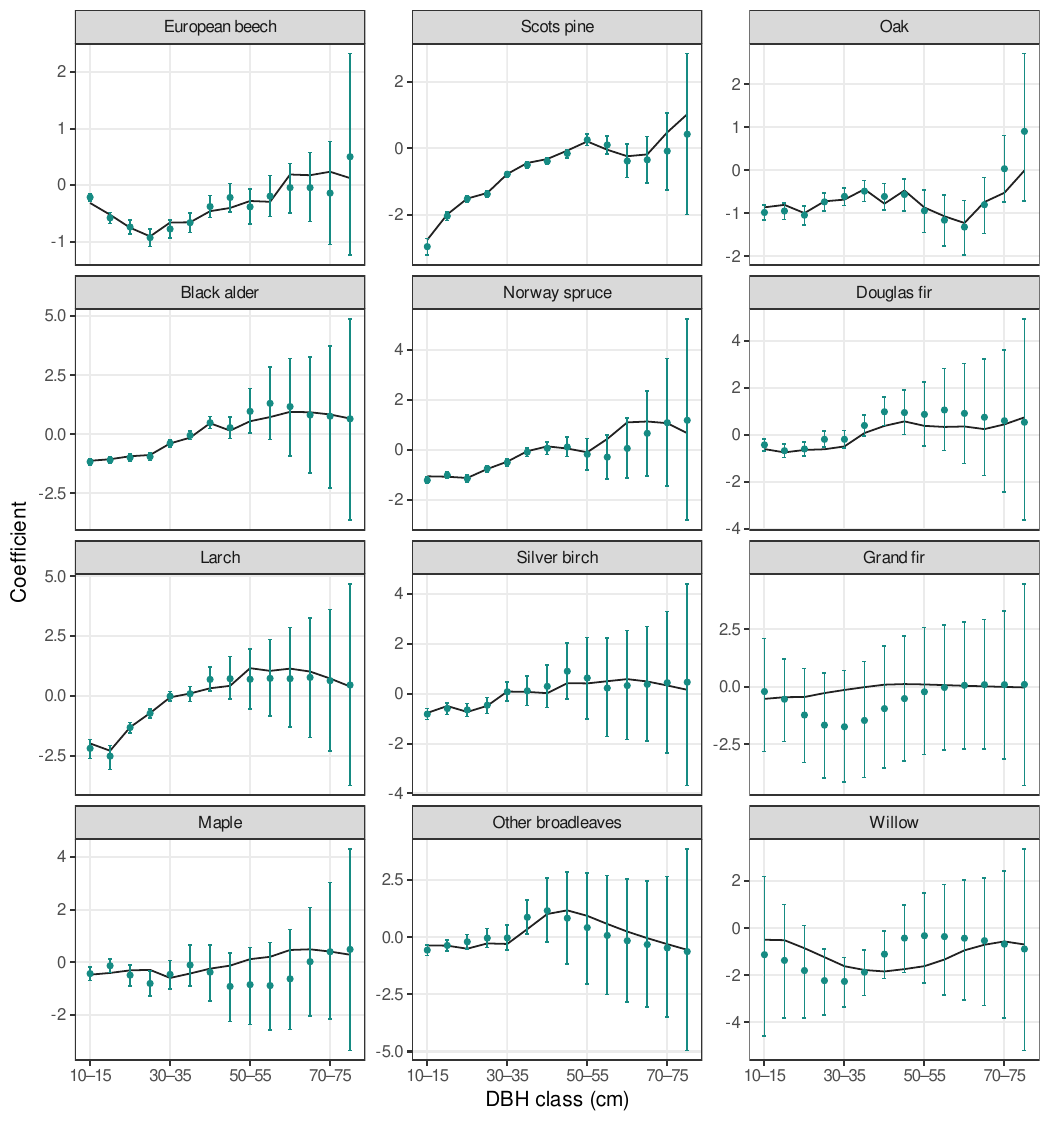}
\caption{Conditional DBH intercept estimates for M8, separated by species group. Points and vertical bars show class-specific posterior means and pointwise 95\% credible intervals; solid lines connect generating coefficients as visual guides. Vertical scales vary by species. The terminal 80+ cm class receives the remaining conditional probability and has no coefficient.}
\label{fig:simulated_m8_dbh_intercept_recovery}
\end{figure}

\begin{figure}[p]
\centering
\includegraphics[width=\textwidth,height=0.73\textheight,keepaspectratio]{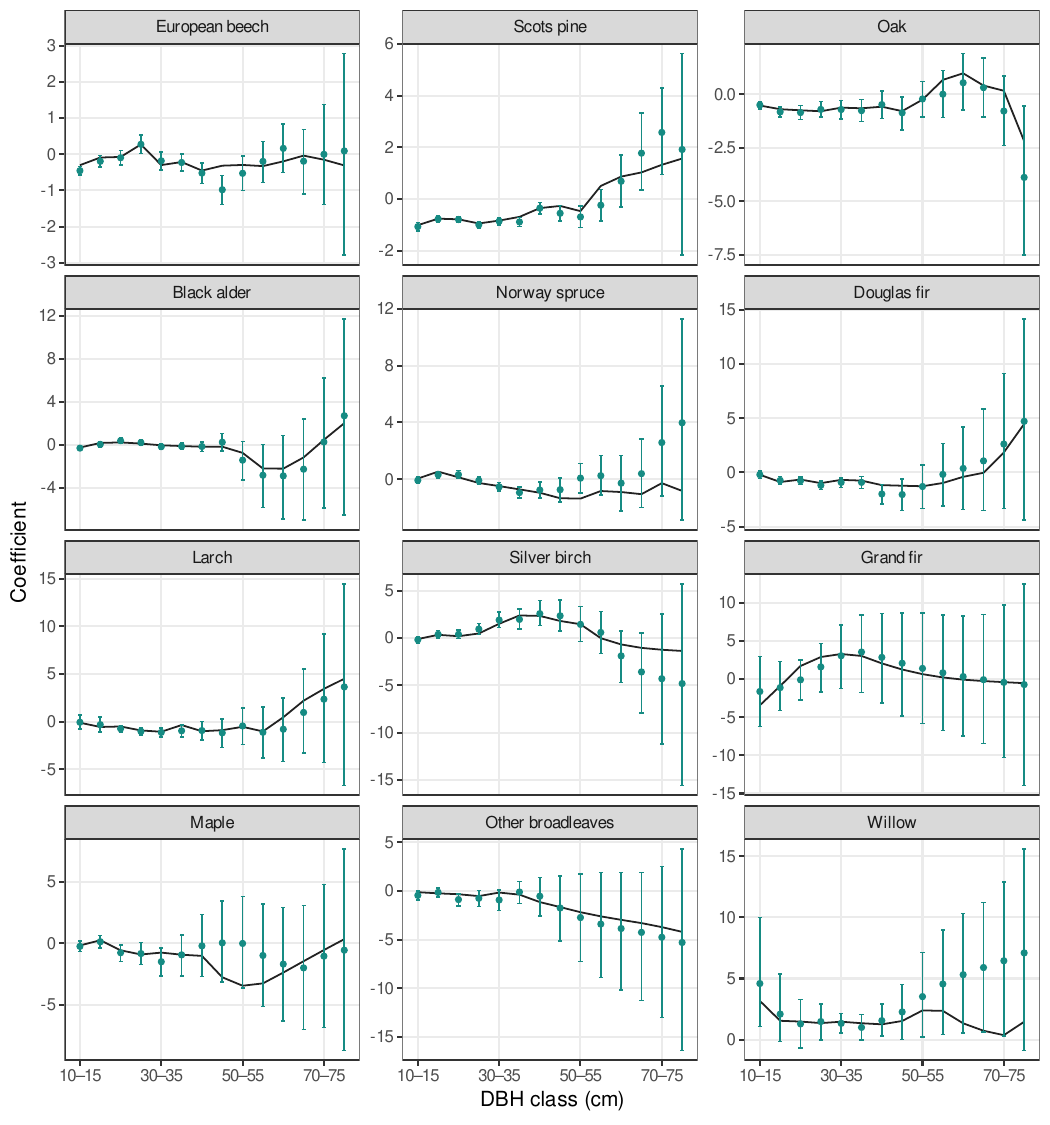}
\caption{Conditional DBH coefficient estimates for the standardized height-p95 covariate in M8. Points and vertical bars show class-specific posterior means and pointwise 95\% credible intervals; solid lines connect generating coefficients as visual guides. Panels separate species groups with individual vertical scales. The terminal 80+ cm class has no coefficient.}
\label{fig:simulated_m8_dbh_p95_recovery}
\end{figure}

\begin{figure}[p]
\centering
\includegraphics[width=\textwidth,height=0.73\textheight,keepaspectratio]{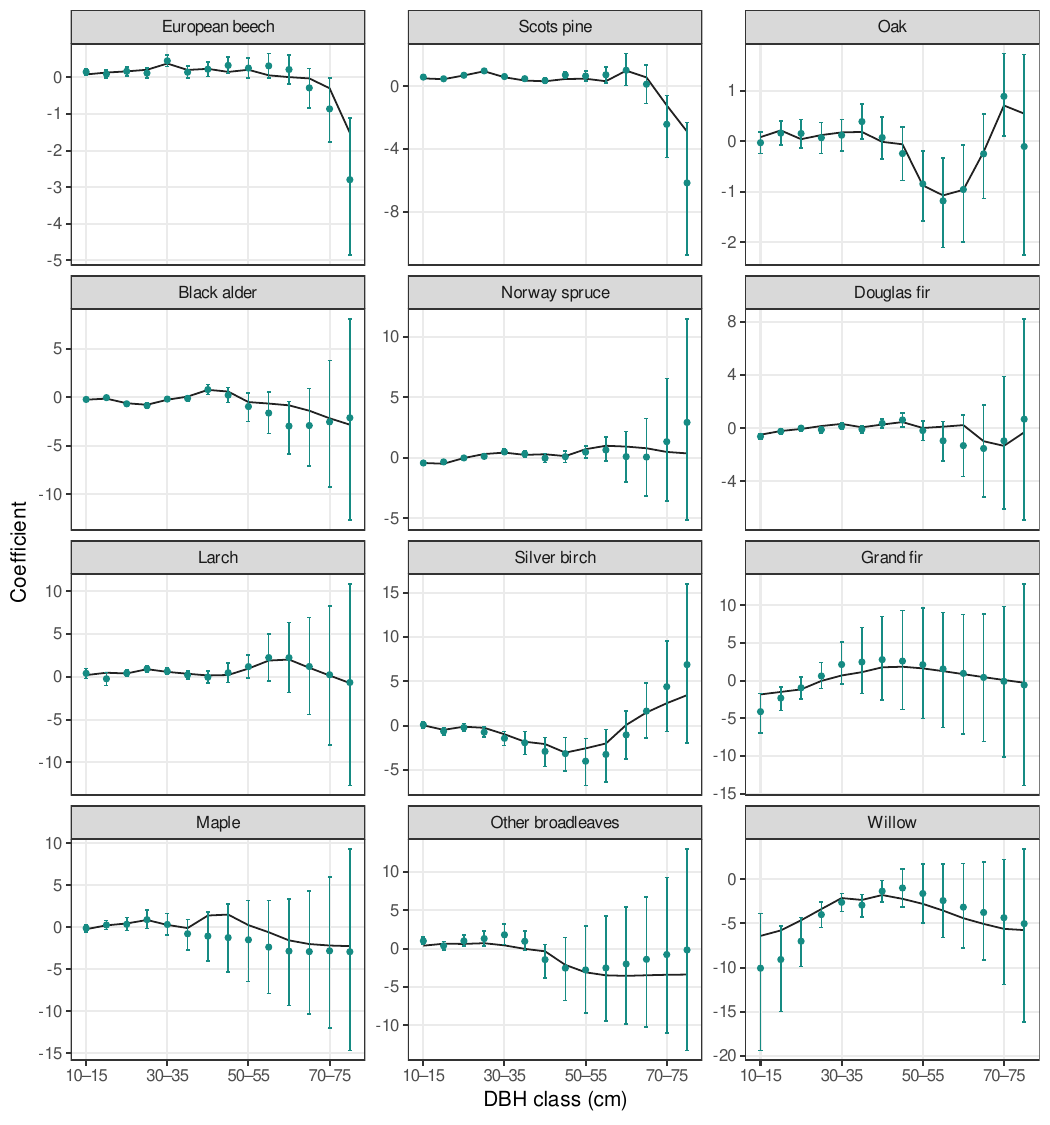}
\caption{Conditional DBH coefficient estimates for the standardized canopy-cover covariate in M8. Points and vertical bars show class-specific posterior means and pointwise 95\% credible intervals; solid lines connect generating coefficients as visual guides. Panels separate species groups with individual vertical scales. The terminal 80+ cm class has no coefficient.}
\label{fig:simulated_m8_dbh_cover_recovery}
\end{figure}

\begin{figure}[p]
\centering
\includegraphics[width=\textwidth,height=0.73\textheight,keepaspectratio]{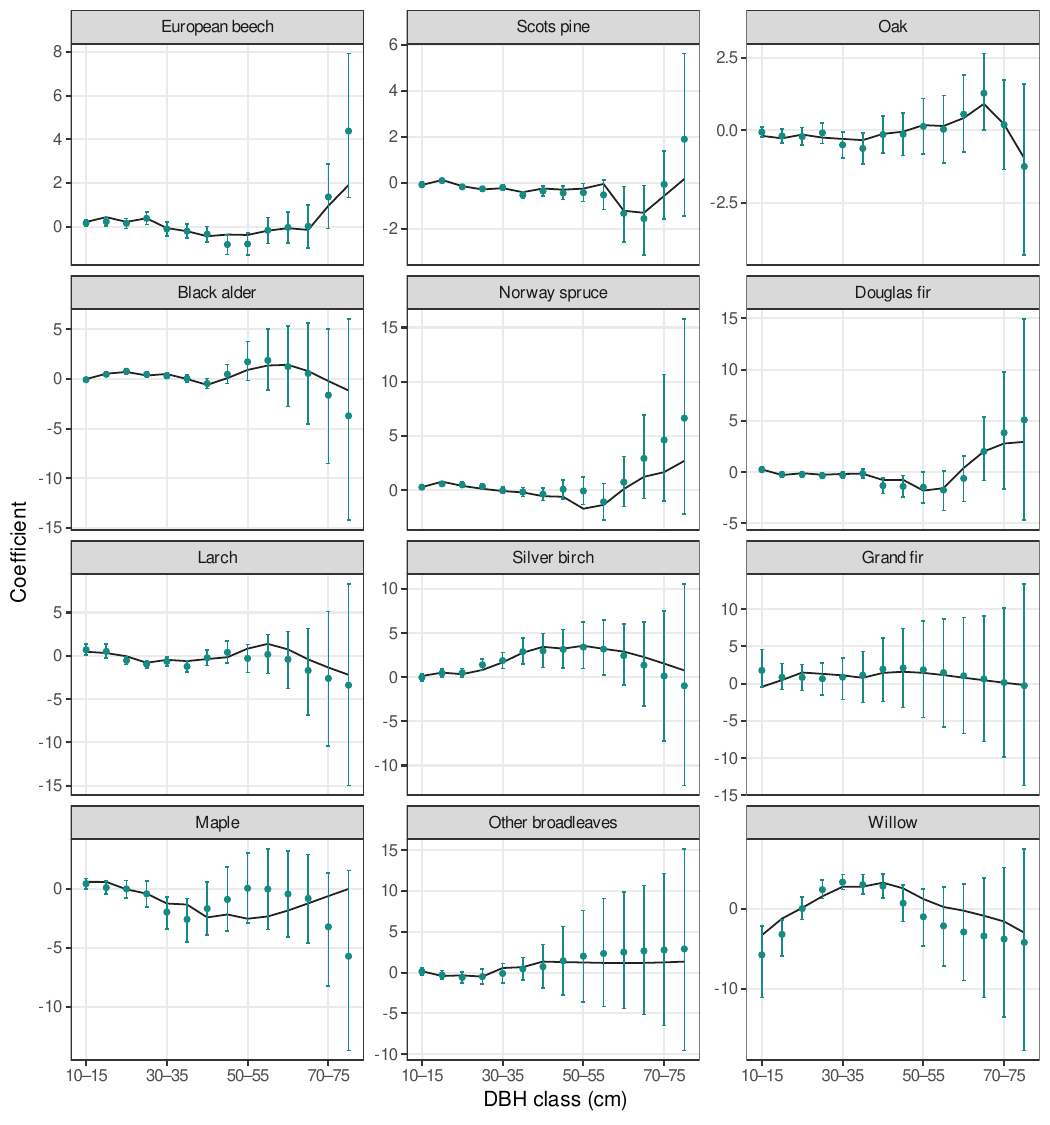}
\caption{Conditional DBH coefficient estimates for the standardized midheight covariate in M8. Points and vertical bars show class-specific posterior means and pointwise 95\% credible intervals; solid lines connect generating coefficients as visual guides. Panels separate species groups with individual vertical scales. The terminal 80+ cm class has no coefficient.}
\label{fig:simulated_m8_dbh_mid_recovery}
\end{figure}

\tabref{tab:simulated_m8_scalar_parameter_recovery} compares true and estimated magnitude coefficients and spatial covariance parameters. \figrefs{fig:simulated_m8_species_coefficient_recovery}{fig:simulated_m8_factor_loading_recovery} show species coefficients and factor loadings; \figrefrange{fig:simulated_m8_dbh_intercept_recovery}{fig:simulated_m8_dbh_mid_recovery} show conditional DBH coefficients. These figures also document the generating parameter values used throughout the repeated-population experiment. Pointwise 95\% credible intervals contain the true values of all four magnitude coefficients, 43 of 44 species coefficients, 651 of 672 conditional DBH coefficients, and all 21 factor loadings.

\tabref{tab:simulated_m8_scalar_parameter_recovery} reports $\widehat R$ and effective sample sizes (ESS). An intercept and its spatial contribution can offset one another in the linear predictor, so the likelihood more directly informs their sum than its separate components. We therefore also assess intercepts combined with their mean spatial contributions over the sampled plots. For magnitude, this quantity is $\beta_0+\bar v$, where $\bar v=n^{-1}\sum_i v(\bs_i)$; it has $\widehat R=1.000$ and ESS 2,930. Analogous combined intercepts are assessed for species composition.

Using these combined intercepts, 94.2\% of $\widehat R$ values are below 1.05. Twenty-eight values exceed 1.1: two first-factor loadings and twenty-six grand fir or willow regression quantities. The maximum is 1.634 for willow's midheight coefficient in the 10--15 cm DBH class. The conditional DBH coefficients with elevated $\widehat R$ concern classes from 10 to 40 cm for grand fir and classes from 10 to 25 cm and 65 to 80 cm for willow. Grand fir occurs on seven plots in this simulated population. These diagnostics indicate slower mixing for some species and size-class regression terms, although the estimated spatial covariance parameters show good agreement between chains.

\begin{figure}[!ht]
\centering
\includegraphics[width=\textwidth,height=0.73\textheight,keepaspectratio]{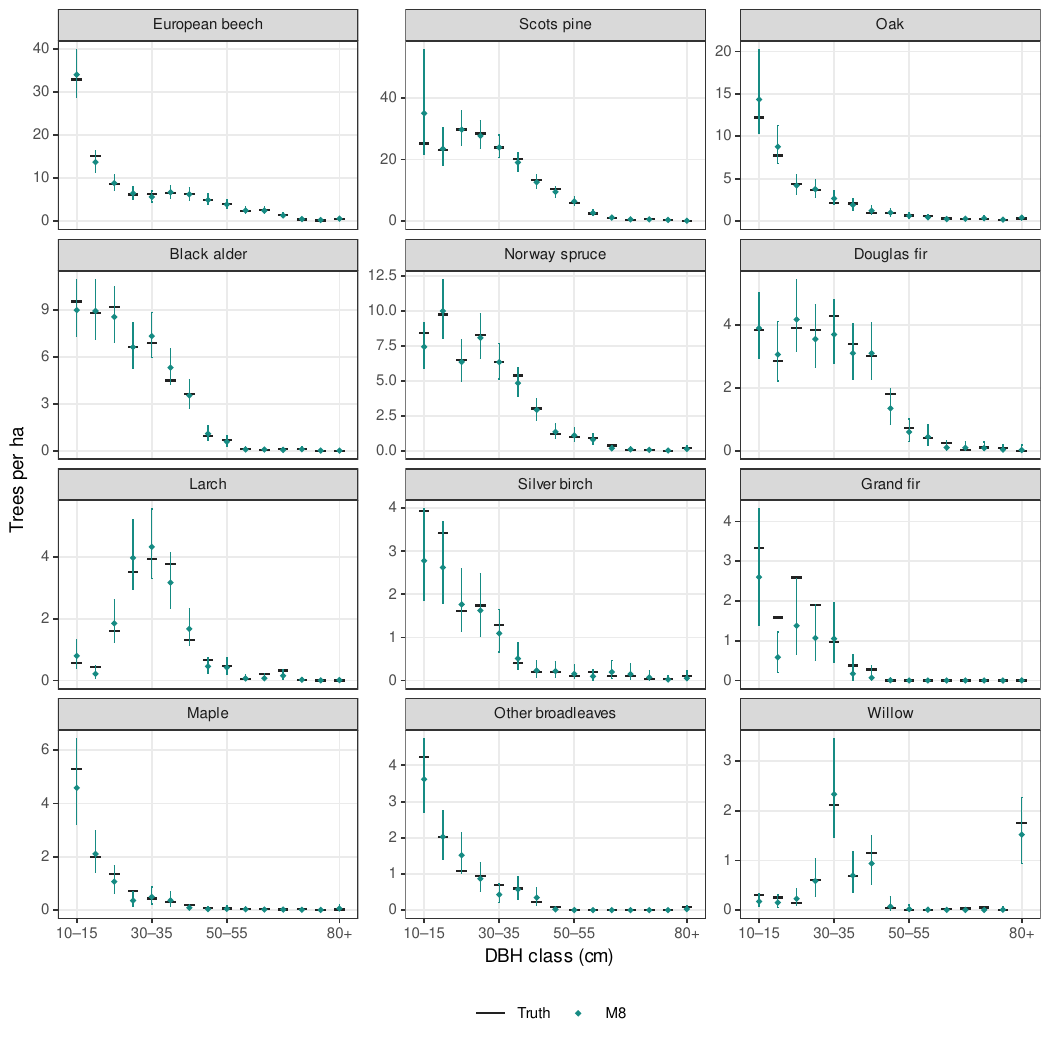}
\caption{Forest species-by-DBH densities in the first simulated population. Horizontal black bars show true densities; colored points and vertical bars show M8 posterior means and pointwise 95\% predictive intervals. Vertical scales vary by species; the terminal class includes all diameters at least 80 cm.}
\label{fig:simulated_m8_forest_species_dbh_estimates}
\end{figure}

\figref{fig:simulated_m8_forest_species_dbh_estimates} compares estimated and true forest species-by-DBH densities; pointwise 95\% predictive intervals contain the true density in 174 of 180 cells (96.7\%).

\begin{figure}[!ht]
\centering
\includegraphics[width=0.9\textwidth]{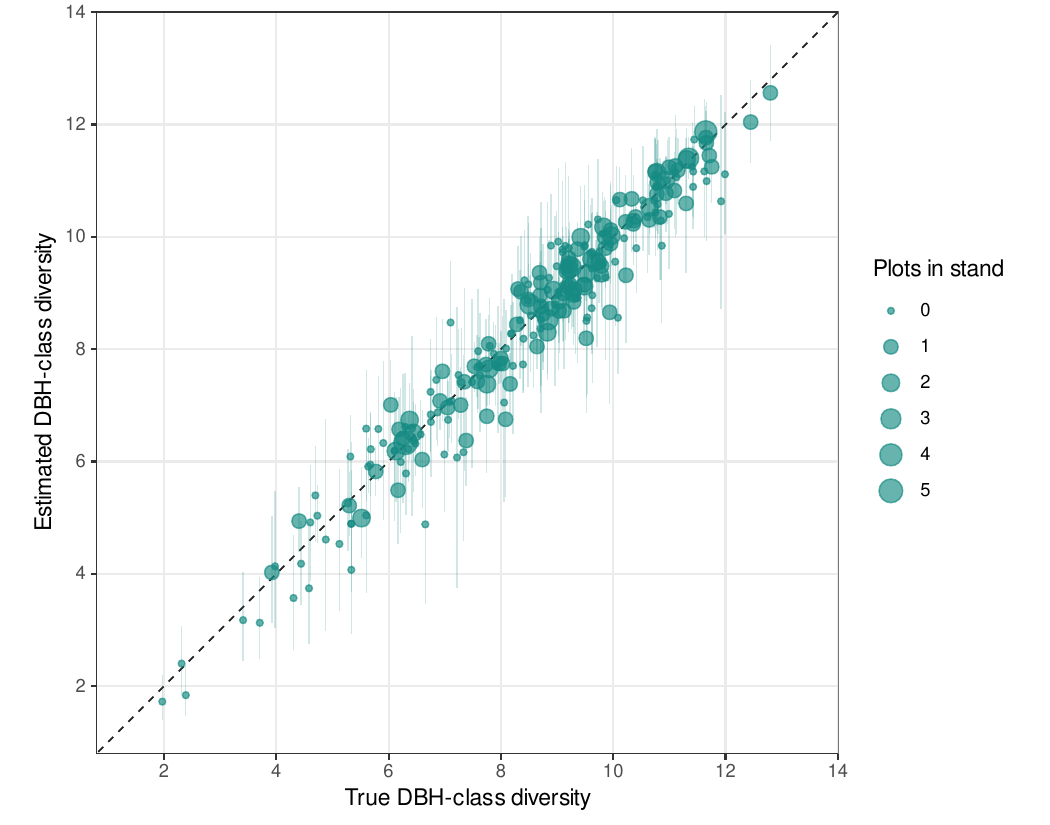}
\caption{Estimated versus true stand DBH-class diversity in the first simulated population. Points show M8 posterior means, with circle size indicating the number of plots in each stand; stands without plots are also shown. Faint vertical bars show pointwise 95\% predictive intervals, and the dashed line marks equality.}
\label{fig:simulated_stand_diversity_recovery}
\end{figure}

\figref{fig:simulated_stand_diversity_recovery} compares estimated and true DBH-class diversity, calculated using \eqnrefs{eq:stand-dbh-proportion}{eq:dbh-diversity}, for all 249 stands, including 139 without sampled plots. Mean error is $-0.076$ effective DBH classes and RMSE is 0.491. Pointwise 95\% predictive intervals contain the true diversity for 233 stands (93.6\%).

\begin{table}[!ht]
\centering
\small
\caption{M8 magnitude regression and spatial covariance parameter estimates for the first simulated population. The combined intercept $\beta_0+\bar v$ includes the mean spatial effect at the 151 plots. Intervals are pointwise 95\% credible intervals; ESS is the effective sample size across the three chains.}
\label{tab:simulated_m8_scalar_parameter_recovery}
\begin{tabular}{@{}lrrrrr@{}}
\toprule
Parameter & True & Posterior mean & 95\% credible interval & $\widehat R$ & ESS \\
\midrule
$\beta_0$ & 5.844 & 5.914 & (5.331, 6.255) & 1.093 & 43 \\
$\beta_0 + \bar v$ & 5.973 & 5.976 & (5.949, 6.004) & 1.000 & 2930 \\
$\beta_{\mathrm{p95}}$ & -0.121 & -0.214 & (-0.337, -0.089) & 1.007 & 578 \\
$\beta_{\mathrm{cover}}$ & 0.193 & 0.205 & (0.090, 0.320) & 1.010 & 445 \\
$\beta_{\mathrm{mid}}$ & 0.123 & 0.003 & (-0.124, 0.133) & 1.006 & 529 \\
$\sigma_v^2$ & 0.500 & 0.502 & (0.320, 0.912) & 1.021 & 599 \\
$\phi_v$ & 3.000 & 2.939 & (1.273, 4.760) & 1.019 & 498 \\
$\phi_{w,1}$ & 2.000 & 1.862 & (0.393, 5.056) & 1.073 & 49 \\
$\phi_{w,2}$ & 5.000 & 5.813 & (2.643, 10.224) & 1.002 & 623 \\
\bottomrule
\end{tabular}
\end{table}
\FloatBarrier

\begin{figure}[p]
\centering
\includegraphics[width=\textwidth]{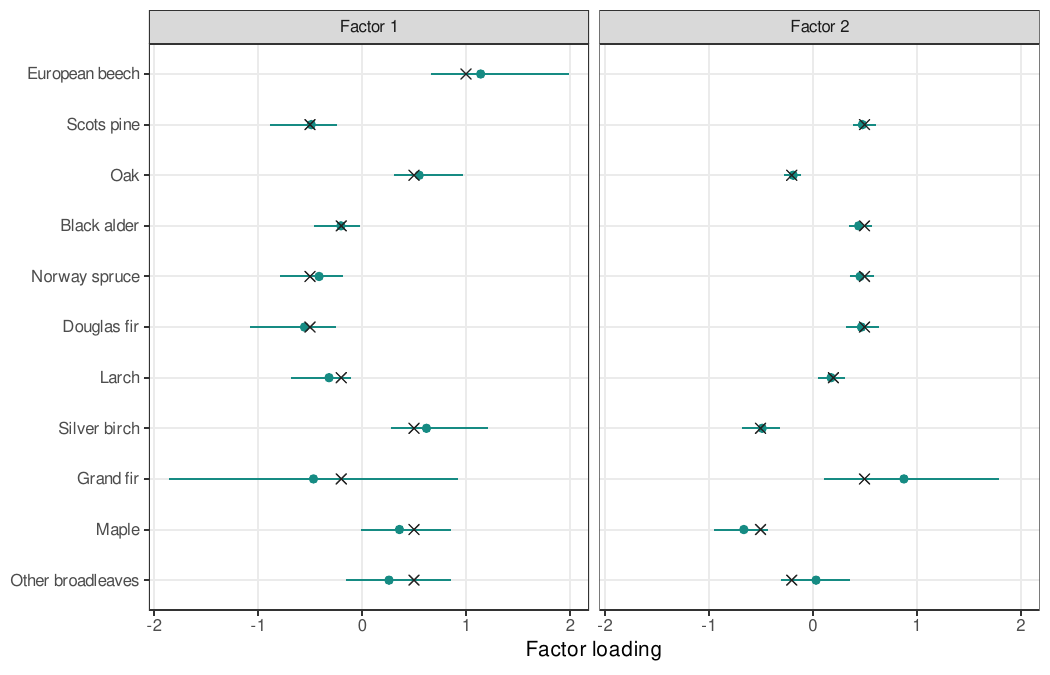}
\caption{Species-composition loading estimates for M8's two spatial factors. Points and horizontal bars show posterior means and pointwise 95\% credible intervals; crosses mark generating values. The European beech loading for factor 1 and Scots pine loading for factor 2 are positive estimated anchor diagonals. The European beech loading for factor 2 is fixed at zero and omitted.}
\label{fig:simulated_m8_factor_loading_recovery}
\end{figure}

\FloatBarrier
\section{Supplementary analysis settings and results}
\label{app:application_computation}

\subsection{Prior settings}
\label{sec:application_priors}

The simulation and Weisse Krug analyses use the prior families in \secref{sec:prior_specification} of the main text, with the following settings for components included in each model. The zero-mean normal coefficient priors have variances $\sigma_\beta^2=100$ for magnitude, $\sigma_{\alpha,0}^2=9$ for composition intercepts, and $\sigma_\alpha^2=100$ for composition slopes. Where RW2 smoothing is included, the smoothing variances $\tau^2$ in \eqnref{eq:smoothing-prior} are fixed at 1 for conditional DBH intercepts and 4 for conditional DBH slopes.

The magnitude spatial variance has an $\IG(2,1)$ prior, with shape 2 and scale 1. The loading prior variance is $c_\lambda=1$, giving standard normal priors for unrestricted loadings and the corresponding half-normal priors for positive anchor diagonals. Models M1--M5 estimate negative-binomial size using a gamma prior with shape 2 and rate 0.2; M6--M8 fix $r_M=1000$.

Each spatial decay has an independent uniform prior with bounds $\phi_L=3/(2d_{\max})$ and $\phi_U=6/d_{\mathrm{nn}}$, where $d_{\max}$ is the largest distance between fitting locations and $d_{\mathrm{nn}}$ is their median nearest-neighbor distance. With distances measured in kilometers, these bounds are approximately 0.370 and 33.453 for the full set of 151 plots used in both analyses. The bounds are recalculated from the training locations for each cross-validation fit.

\subsection{MCMC settings and diagnostics}
\label{sec:application_mcmc}

\tabref{tab:application_mcmc_settings} records the MCMC settings for the simulation and Weisse Krug analyses. Simulation cross-validation is limited to the first population. Thinning is applied after discarding burn-in; retained draws are pooled across chains where applicable.

\begin{table}[!ht]
\centering
\footnotesize
\setlength{\tabcolsep}{3pt}
\caption{MCMC settings. Iterations and burn-in are per chain; retained draws are totals across chains.}
\label{tab:application_mcmc_settings}
\begin{tabular*}{\textwidth}{@{\extracolsep{\fill}}lrrrrr@{}}
\toprule
Analysis & Chains & Iterations & Burn-in & Thinning & \shortstack{Retained\\draws} \\
\midrule
\shortstack[l]{Simulation: M8\\per population} & 1 & 25,000 & 10,000 & 10 & 1,500 \\
\shortstack[l]{Simulation: each\\cross-validation fit} & 1 & 50,000 & 20,000 & 10 & 3,000 \\
\shortstack[l]{Weisse Krug: each\\cross-validation fit} & 1 & 50,000 & 20,000 & 10 & 3,000 \\
Weisse Krug: M7 and M8, all plots & 3 & 50,000 & 20,000 & 10 & 9,000 \\
\bottomrule
\end{tabular*}
\end{table}

For each simulated population, parameter summaries use all 1,500 retained draws. Forest and stand prediction uses every third retained draw, giving 500 joint prediction draws. The first population uses the first chain of the multichain fit described in \secref{sec:simulation_generating_parameters}, with the same 500-draw prediction procedure. Each cross-validation fit uses all 3,000 retained draws for assessment. The all-plot Weisse Krug M7 fit uses all 9,000 draws for parameter and prediction summaries. We also fit M8 with these multichain settings to assess its parameter convergence.

The repeated-population and cross-validation fits use single chains, so between-chain $\widehat R$ is unavailable.

\subsection{Predictive performance across density targets}
\label{sec:real_cv_targets}

Using the same leave-one-plot-out predictions as \tabref{tab:real_cv_species_dbh}, we assess overall, species, DBH-class, and species-by-DBH densities. Marginal densities are obtained by summing the appropriate categories within each posterior draw before computing CRPS as described in \secref{sec:cv_metrics}. \tabref{tab:real_cv_crps_targets} compares the candidate models across these targets. M7 and M8 have the lowest CRPS for species and DBH-class densities, while M6--M8 perform similarly for overall density, consistent with their shared magnitude-model structure.

\begin{figure}[p]
\centering
\includegraphics[width=\textwidth,height=0.82\textheight,keepaspectratio]{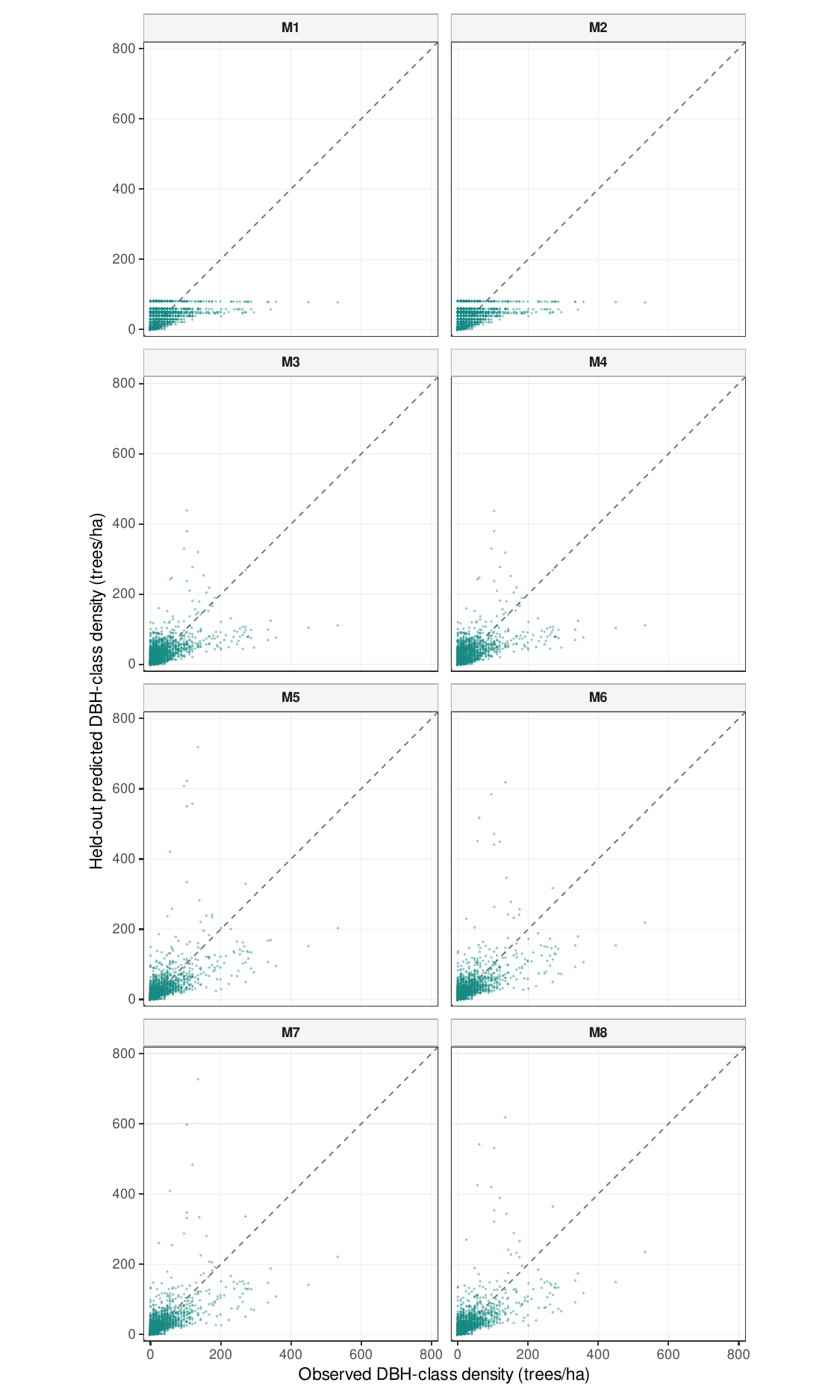}
\caption{Observed versus held-out predicted DBH-class density for each candidate model. Points show posterior predictive means from leave-one-plot-out fits, with densities summed across species within each DBH class. Each panel contains 2,265 plot-by-class pairs (151 plots $\times$ 15 classes). The dashed line marks equality, and all panels use the same axis limits.}
\label{fig:real_cv_dbh}
\end{figure}

\figref{fig:real_cv_dbh} compares posterior predictive means with observed DBH-class densities at held-out plots. Summing across species assesses prediction of size structure irrespective of species identity. DBH-class densities predicted over the forest grid are aggregated to stands to obtain the diversity summaries in \figref{fig:real_stand_diversity}.

\begin{table}[!ht]
\centering
\small
\caption{Leave-one-plot-out CRPS for the 151 Weisse Krug plots, averaged over plots and the categories defining each target. All scores are in trees/ha; lower values indicate better predictive performance. Compare models within each column.}
\label{tab:real_cv_crps_targets}
\begin{tabular*}{\textwidth}{@{\extracolsep{\fill}}lrrrr@{}}
\toprule
Model & \multicolumn{1}{c}{\shortstack{Overall\\density}} & \multicolumn{1}{c}{\shortstack{Species\\density}} & \multicolumn{1}{c}{\shortstack{DBH-class\\density}} & \multicolumn{1}{c@{}}{\shortstack{Species-by-DBH\\density}} \\
\midrule
M1 & 111.145 & 30.438 & 14.305 & 2.044 \\
M2 & 111.008 & 30.442 & 14.295 & 2.042 \\
M3 & 92.189 & 33.081 & 13.877 & 2.155 \\
M4 & 91.945 & 33.082 & 13.855 & 2.154 \\
M5 & 92.305 & 28.028 & 12.319 & 1.932 \\
M6 & 89.217 & 27.845 & 12.121 & 1.925 \\
M7 & 89.593 & 22.756 & 11.812 & 1.742 \\
M8 & 89.391 & 22.485 & 11.908 & 1.727 \\
\bottomrule
\end{tabular*}
\end{table}

\FloatBarrier

\subsection{M7 parameter estimates for the Weisse Krug analysis}
\label{sec:real_parameter_estimates}

We summarize the M7 fit to all 151 plots using the 9,000 retained draws from three chains. Prior and MCMC settings are given in \secref{sec:application_priors} and \secref{sec:application_mcmc}. The figures below parallel the parameter summaries for the first simulated population in \secref{sec:simulation_generating_parameters}.

\tabref{tab:real_scalar_parameters} reports the magnitude regression coefficients, magnitude spatial variance, and magnitude and composition spatial decays, together with $\widehat R$ and ESS. As in the simulation analysis, we also assess intercepts combined with their mean spatial contributions over the sampled plots because these components can offset one another in the linear predictor. The magnitude combined intercept $\beta_0+\bar v$ has $\widehat R=1.000$ and ESS 6,578. Species combined intercepts have $\widehat R<1.1$ except for black alder ($\widehat R=1.153$).

Using the combined intercepts, 97.4\% of $\widehat R$ values are below 1.05. The eleven values above 1.1 are primarily grand fir regression quantities. The maximum is 1.514 for grand fir's species midheight coefficient; the affected conditional coefficients span classes from 10 to 20 cm. Grand fir occurs on only two plots, both in the same stand with similar covariate values, limiting the information available for separating its regression effects (\tabref{tab:real_inventory_species}). Its regression chains mix slowly, whereas magnitude, spatial covariance, and factor-loading parameters show good agreement between chains. M8 shows greater loading-chain disagreement: 20 of its 21 loadings have $\widehat R>1.1$, with a maximum of 5.316 and the largest values associated with other broadleaves, oak, black alder, and Norway spruce. Its small predictive gain and poorer mixing motivated selection of M7.

\begin{table}[!ht]
\centering
\small
\caption{M7 magnitude regression and spatial covariance parameter estimates for the Weisse Krug analysis. The combined intercept $\beta_0+\bar v$ includes the mean spatial effect at the 151 plots. Intervals are pointwise 95\% credible intervals; ESS is the effective sample size across the three chains.}
\label{tab:real_scalar_parameters}
\begin{tabular}{@{}lrrrr@{}}
\toprule
Parameter & Posterior mean & 95\% credible interval & $\widehat R$ & ESS \\
\midrule
$\beta_0$ & 5.756 & (5.658, 5.854) & 1.001 & 903.1 \\
$\beta_0 + \bar v$ & 5.753 & (5.724, 5.781) & 1.000 & 6578.0 \\
$\beta_{\mathrm{p95}}$ & -0.068 & (-0.176, 0.040) & 1.000 & 1656.7 \\
$\beta_{\mathrm{cover}}$ & 0.236 & (0.145, 0.329) & 1.001 & 1387.0 \\
$\beta_{\mathrm{mid}}$ & 0.213 & (0.101, 0.328) & 1.000 & 1508.9 \\
$\sigma_v^2$ & 0.175 & (0.131, 0.233) & 1.000 & 5534.1 \\
$\phi_v$ & 12.197 & (5.397, 30.554) & 1.001 & 2884.3 \\
$\phi_{w,1}$ & 17.809 & (9.040, 32.321) & 1.001 & 4031.2 \\
\bottomrule
\end{tabular}
\end{table}
\FloatBarrier

\begin{figure}[p]
\centering
\includegraphics[width=\textwidth]{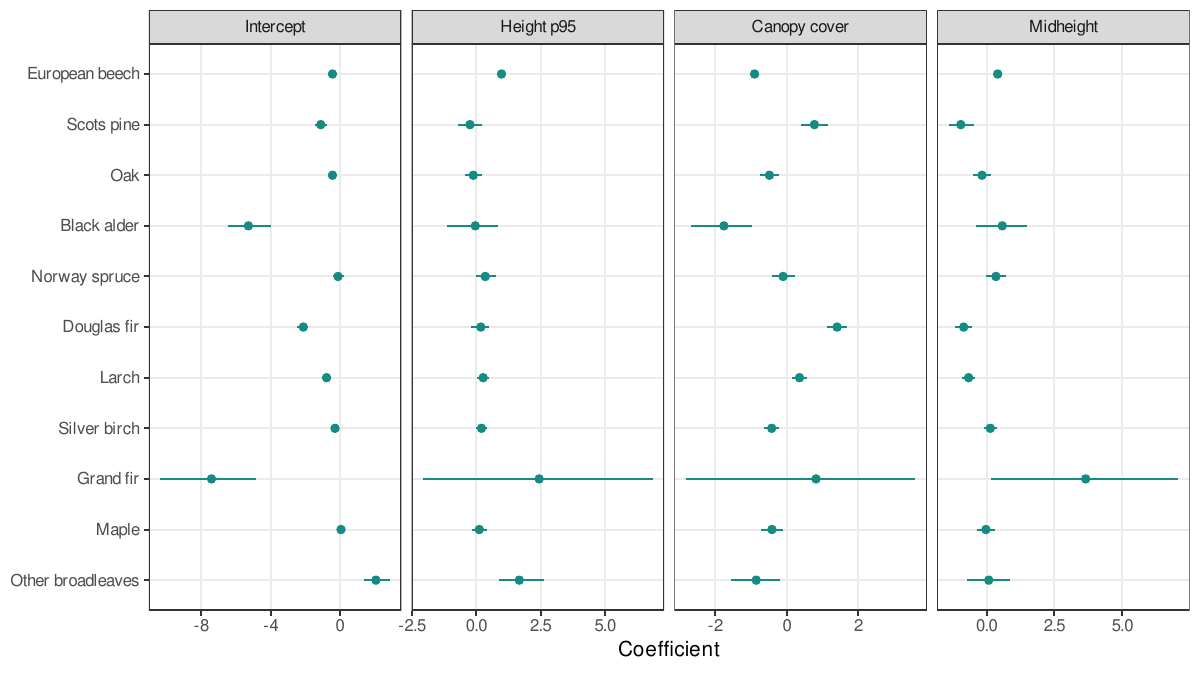}
\caption{M7 species-composition regression coefficient estimates for Weisse Krug. Points and horizontal bars show posterior means and pointwise 95\% credible intervals. Panels separate intercepts and the three standardized covariate coefficients, with horizontal scales varying by panel. The terminal species group has no coefficient row.}
\label{fig:real_species_coefficients}
\end{figure}

\begin{figure}[p]
\centering
\includegraphics[width=\textwidth,height=0.73\textheight,keepaspectratio]{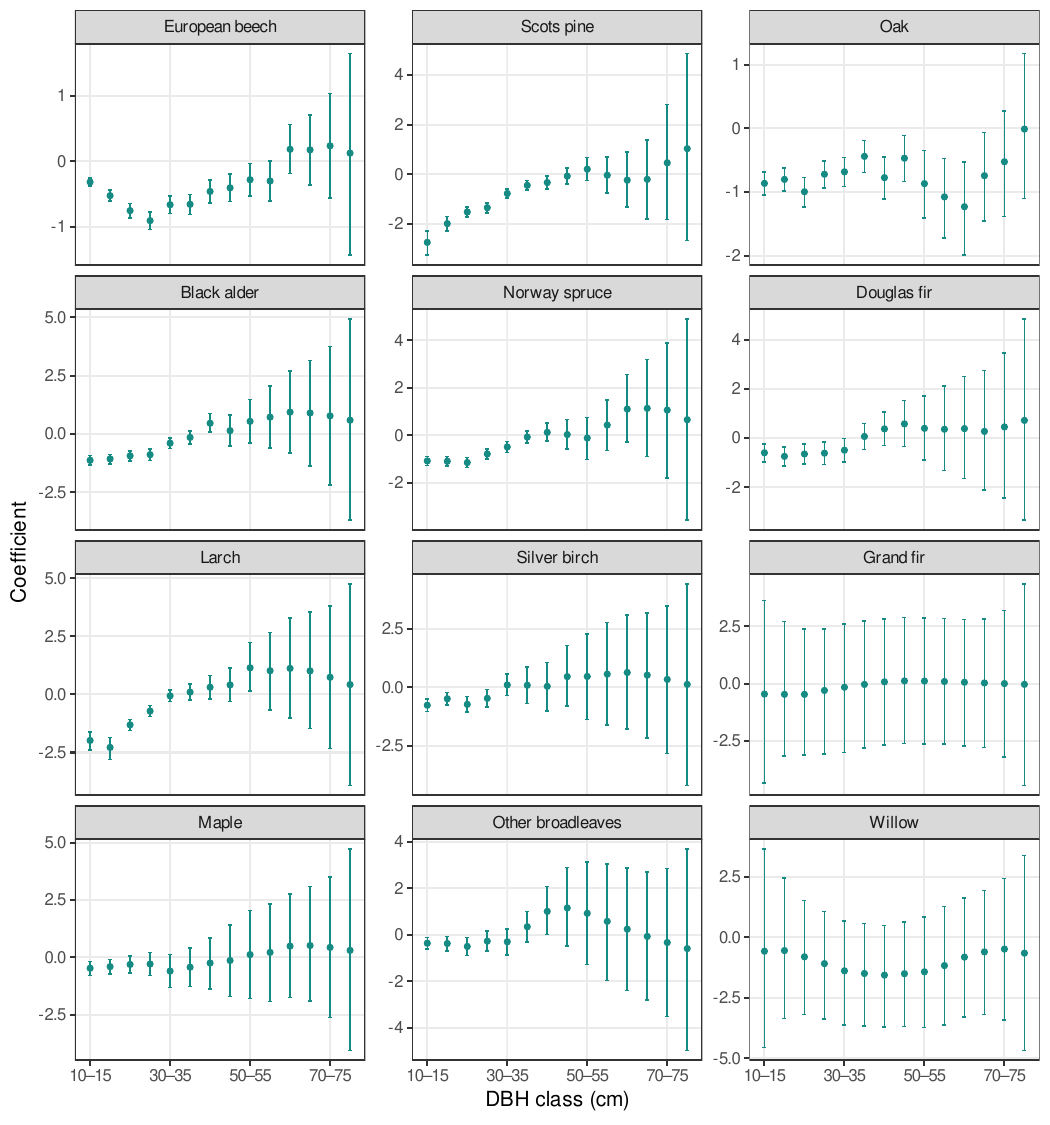}
\caption{M7 conditional DBH intercept estimates for Weisse Krug, separated by species group. Points and vertical bars show class-specific posterior means and pointwise 95\% credible intervals. Vertical scales vary by species. The terminal 80+ cm class receives the remaining conditional probability and has no coefficient.}
\label{fig:real_dbh_intercept}
\end{figure}

\begin{figure}[p]
\centering
\includegraphics[width=\textwidth,height=0.73\textheight,keepaspectratio]{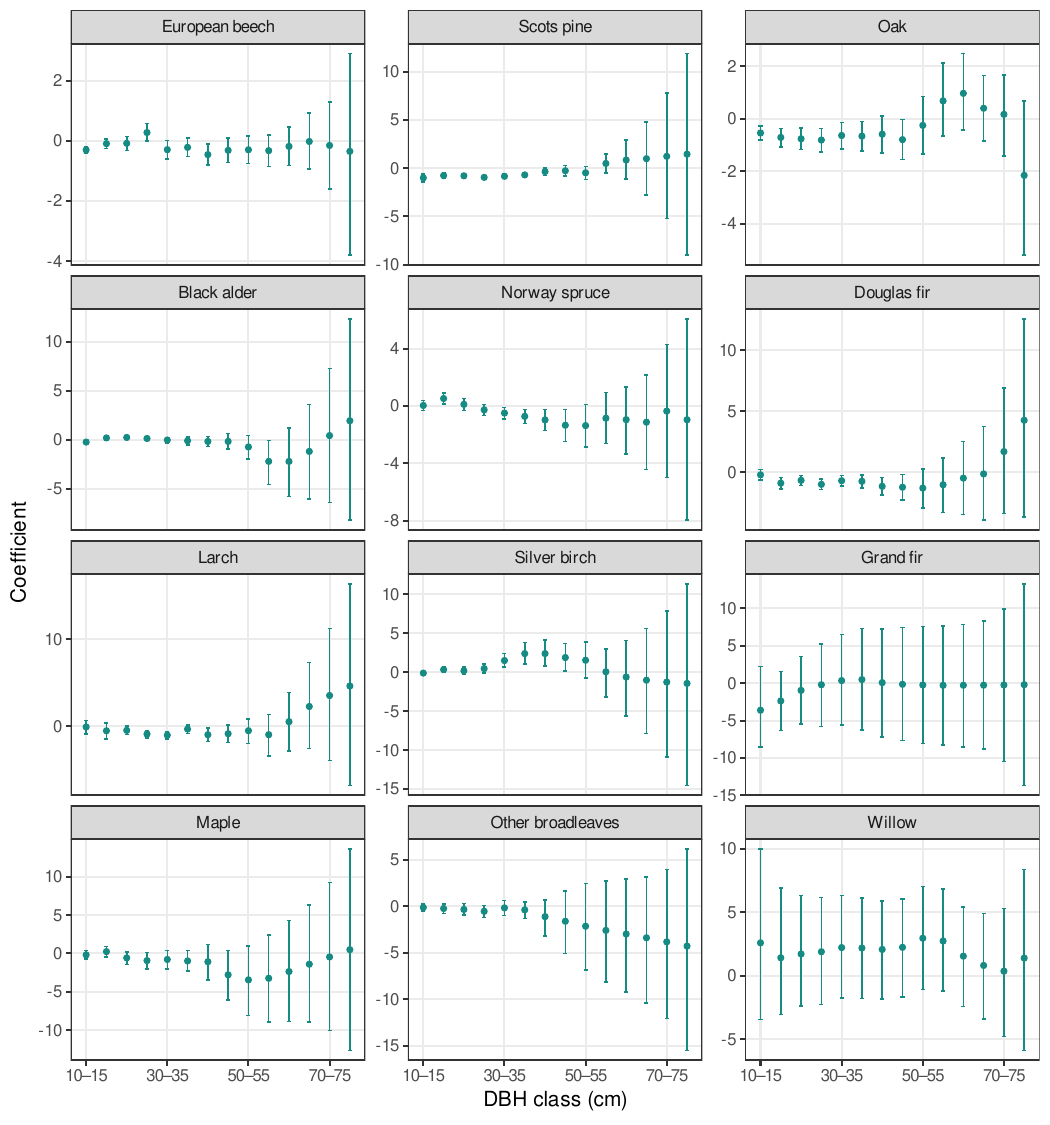}
\caption{M7 conditional DBH coefficient estimates for the standardized height-p95 covariate in the Weisse Krug analysis. Points and vertical bars show class-specific posterior means and pointwise 95\% credible intervals. Panels separate species groups with individual vertical scales. The terminal 80+ cm class has no coefficient.}
\label{fig:real_dbh_p95}
\end{figure}

\begin{figure}[p]
\centering
\includegraphics[width=\textwidth,height=0.73\textheight,keepaspectratio]{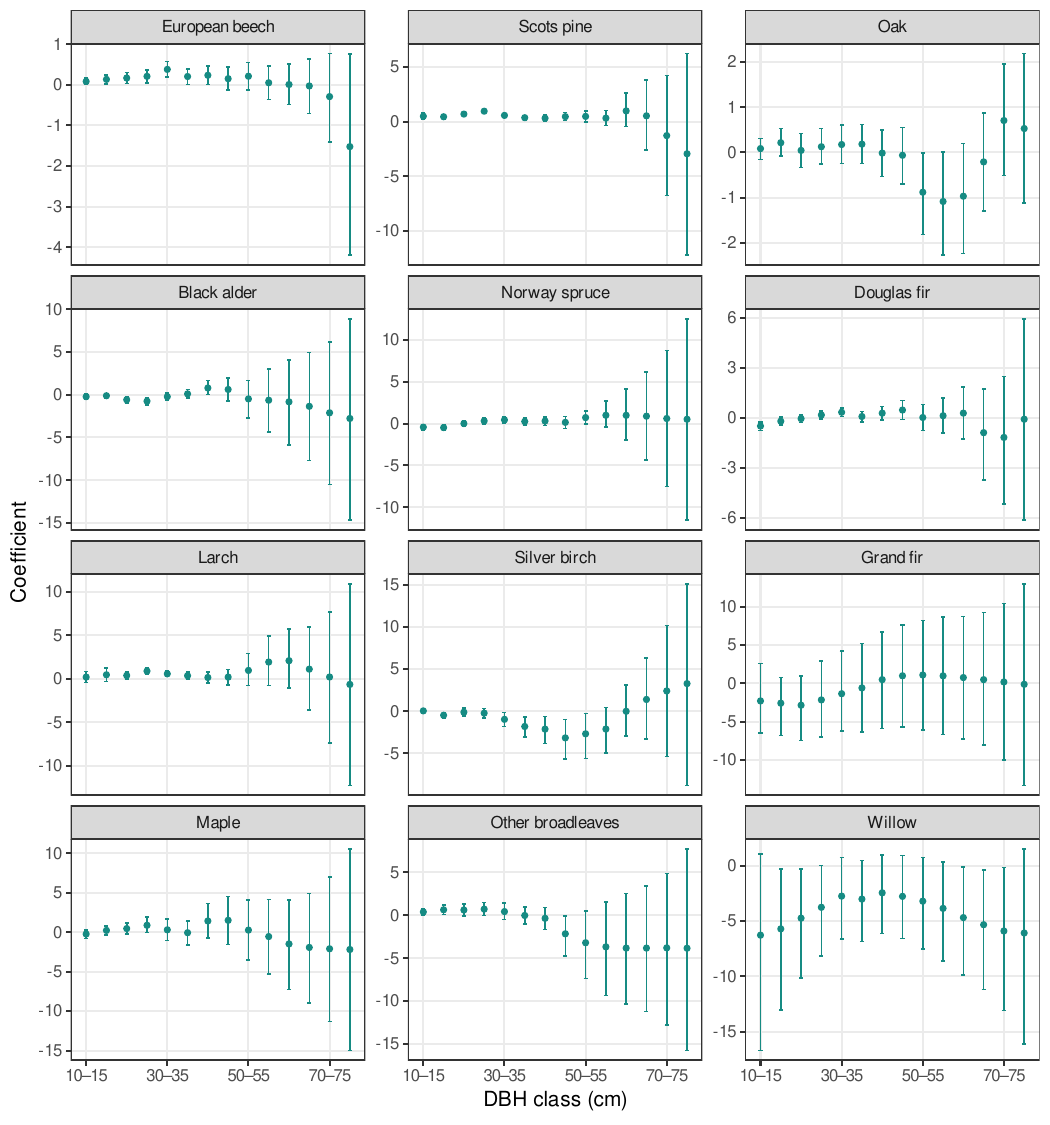}
\caption{M7 conditional DBH coefficient estimates for the standardized canopy-cover covariate in the Weisse Krug analysis. Points and vertical bars show class-specific posterior means and pointwise 95\% credible intervals. Panels separate species groups with individual vertical scales. The terminal 80+ cm class has no coefficient.}
\label{fig:real_dbh_cover}
\end{figure}

\begin{figure}[p]
\centering
\includegraphics[width=\textwidth,height=0.73\textheight,keepaspectratio]{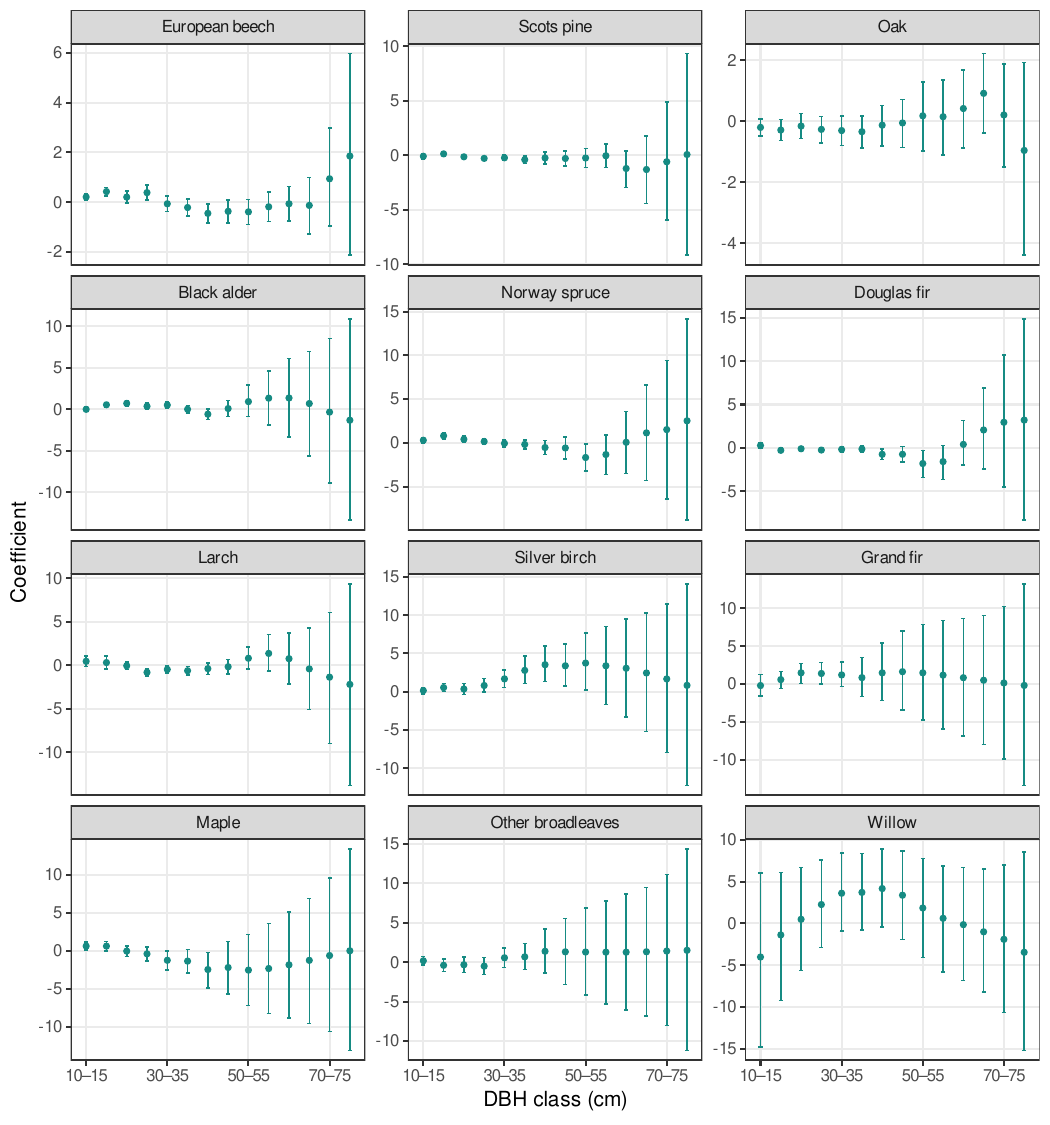}
\caption{M7 conditional DBH coefficient estimates for the standardized midheight covariate in the Weisse Krug analysis. Points and vertical bars show class-specific posterior means and pointwise 95\% credible intervals. Panels separate species groups with individual vertical scales. The terminal 80+ cm class has no coefficient.}
\label{fig:real_dbh_mid}
\end{figure}

\figrefs{fig:real_species_coefficients}{fig:real_factor_loadings} show species regression coefficients and loadings for M7's single spatial composition factor. \figrefrange{fig:real_dbh_intercept}{fig:real_dbh_mid} show conditional DBH intercepts and slopes. Intercepts are displayed in their original, uncentered parameterization.

\begin{figure}[p]
\centering
\includegraphics[width=\textwidth]{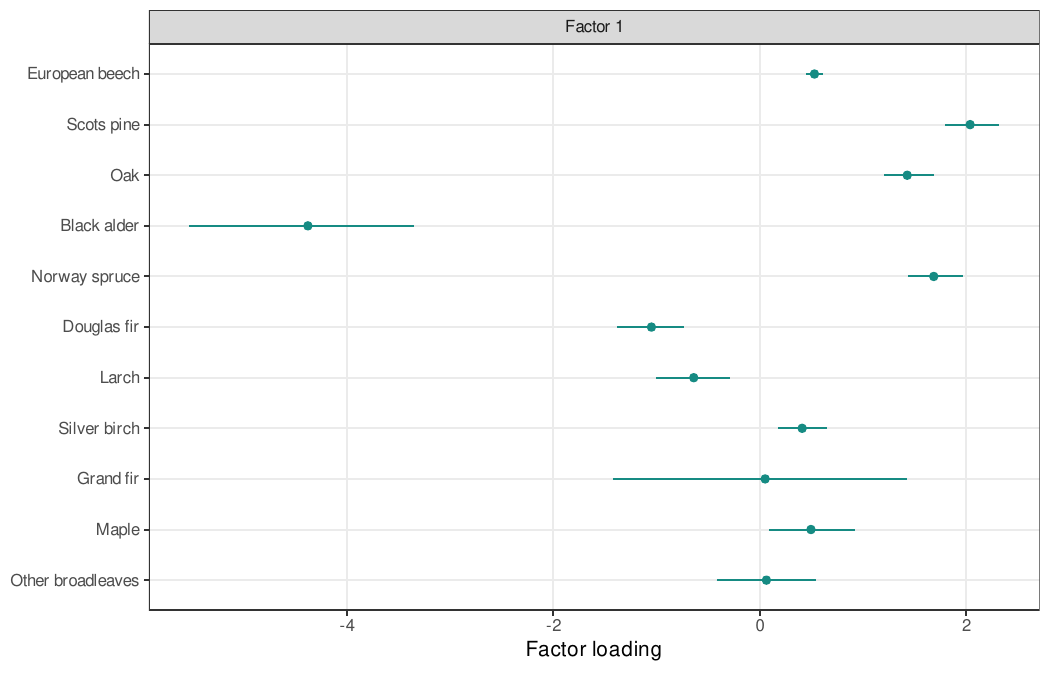}
\caption{Species-composition loading estimates for M7's single spatial factor in the Weisse Krug analysis. Points and horizontal bars show posterior means and pointwise 95\% credible intervals. The European beech loading is the positive estimated anchor diagonal.}
\label{fig:real_factor_loadings}
\end{figure}

\FloatBarrier
\section{Extension to horizontal point sampling}\label{sec:HPS}

The magnitude--composition formulation can be extended to horizontal point sampling, with magnitude defined as basal area per unit area. The formulation below does not explicitly account for boundary effects, which require additional consideration for stand and forest inference (\secref{sec:point_boundary_inference}).

\subsection{Observation model}

Let $m_i$ denote the number of ``in'' trees tallied at point $i$, and let $\BAF_i$ denote its basal area factor (BAF). Each tallied tree contributes $\BAF_i$ units of basal area per unit area to the point-level estimate \citep{grosenbaugh1952plotless,bitterlich1984relascope}, and must be recorded with enough information to assign it to the desired categories. We use this relationship to specify a count model with mean $M_i^{BA}/\BAF_i$, where $M_i^{BA}$ is the expected point-level basal-area summary, expressed per unit area. Specifically, we assume
\begin{equation}
  m_i \given M_i^{BA},r_M^{BA} \sim \NB\left(\frac{M_i^{BA}}{\BAF_i},r_M^{BA}\right),
  \label{eq:point-total}
\end{equation}
where $r_M^{BA}>0$ is the negative-binomial size parameter.

Given the total tally, we model its allocation among categories as
\begin{equation}
  \by_i^{BA}\given m_i,\bp_i^{BA} \sim \Mult(m_i,\bp_i^{BA}),
  \label{eq:point-composition}
\end{equation}
where $\by_i^{BA}=(y_{i1}^{BA},\ldots,y_{iJ}^{BA})\tr$ and $y_{ij}^{BA}$ is the number of tallied trees assigned to category $j$. Because each tallied tree contributes the same BAF, $\bp_i^{BA}$ describes basal-area composition. The expected basal-area density in category $j$ is therefore $M_i^{BA}p_{ij}^{BA}$. The magnitude and composition models in \secref{sec:model} of the main text can be used to estimate these quantities, including species-specific basal-area distributions across DBH classes.

For a tallied tree $t$ at point $i$ with basal area $\mathrm{BA}_{it}$, the tree factor $\mathrm{TF}_{it}=\BAF_i/\mathrm{BA}_{it}$ gives the number of trees per unit area that it represents. Summing tree factors gives the point-level estimate of stem density. Multiplying each tree factor by the corresponding per-tree volume, biomass, or carbon before summation gives estimates of those quantities per unit area. At the stand-table level, convert basal-area density to stem density by dividing each cell's basal area per unit area by its mean basal area per tree.

\subsection{Prediction and aggregation}

For horizontal point sampling, we predict the total tally and its allocation among categories using \eqnrefs{eq:point-total}{eq:point-composition}, with a specified BAF at each prediction location. Multiplying each predicted category tally by that location's BAF gives a posterior predictive basal-area density. Using these densities in \eqnrefs{eq:stand-total}{eq:stand-density}, with the same represented land areas $A(\bs_0)$, gives stand or forest basal-area totals and densities.

\subsection{Boundary considerations}\label{sec:point_boundary_inference}

For horizontal point sampling, \eqnrefs{eq:point-total}{eq:point-composition} provide a working likelihood for total tallies and their allocation among categories, with BAF converting tallies to basal-area summaries. Away from boundaries, interpreting these summaries as local forest quantities assumes that the modeled density and composition adequately represent the regions from which trees can be selected. At points near or outside the forest boundary, restricting tallies to forest trees can reduce expected counts even when forest basal-area density is unchanged. Because inclusion-zone sizes depend on tree DBH, this reduction differs across size categories and cannot generally be represented by a simple spatial area reduction. The current point-sampling model uses BAF to relate tallies to basal-area density, but does not explicitly accommodate the truncation of inclusion zones at forest boundaries.

Estimating stand or forest basal area from these point samples requires accounting for the reduction in expected tallies caused by the boundary. Design-based approaches include mirage and walkthrough corrections and sampling a surrounding buffer that encompasses complete inclusion zones \citep{ducey2004walkthrough}. While design-based corrections ensure unbiased mean estimation, adjusting tallies in this way violates the distributional assumptions of the proposed count likelihoods. Models that explicitly integrate inclusion-zone geometry into the likelihood construction, together with corresponding prediction and aggregation methods, remain an area for further work.

\end{document}